\documentclass[article]{elsarticle}

\usepackage{graphicx}
\usepackage{longtable}
\usepackage{braket}
\usepackage{tabularx}
\usepackage{enumerate}
\usepackage{url}
\usepackage{float}
\usepackage{xspace}
\usepackage{booktabs}
\def\ju{\textit{jun45pn}}
\def\gle{\textit{glekpn}}
\def\gwb{\textit{gwbxg}}
\def\sn{\textit{snt}}

\def\etal{\textit{et al.}}
\usepackage{rotating}
\usepackage{color}
\usepackage{comment}
\usepackage{subcaption}

\begin{document}

\begin{frontmatter}

\title{Cross-shell excitations and the evolution of high spin states in $^{89}\mathrm{Y}$: A shell-model  revisit}
\author[label1]{Sangeeta Das\corref{cor1}\fnref{equal}}
\cortext[cor1]{Corresponding Author}
\ead{dsngeta@gmail.com}
\author[label1,label2]{Arkabrata Gupta\fnref{equal}}
%
\vspace{2cm}
\address[label1]{Institute of Engineering \& Management, Kolkata, University of Engineering and Management, Kolkata, 700160, India}
\address[label2]{Centre of Excellence in Astronomical Studies, Institute of Engineering \& Management, Kolkata, University of Engineering and Management, Kolkata, 700160, India}

\fntext[equal]{Both the authors contributed equally to this work.}

\begin{abstract}

Nuclei in the vicinity of $Z=38$--40 and $N=50$ provide an important testing ground for studying the evolution of nuclear structure across the proton subshell and neutron shell closure. In particular, $^{89}$Y offers an opportunity to investigate the inert core behaviour and effect of particle-hole excitations in generating the observed high spin states. In this work, we perform a systematic large-scale shell-model study of $^{89}$Y using different model spaces and well accepted effective interactions. Excitation energies, electromagnetic transition probabilities, and nucleon occupation probabilities are calculated and compared with available experimental data to explain the microscopic structure of the observed high spin states. The role of neutron excitations across the $N=50$ shell gap is examined in detail. A comparative analysis of the employed effective interactions is performed to assess their applicability in describing the spectroscopic properties of $^{89}$Y.

\end{abstract}


\end{frontmatter}

\date{\today}

\section{\label{Introduction} Introduction}
The mass region around A $\sim$ 85 has attracted sustained theoretical and experimental interest because of the remarkable diversity of nuclear structure phenomena it exhibits. Nuclei in this region display a wide range of intriguing features, including shape coexistence~\cite{shape1}, signature inversion \cite{Sig_sys1}, shear mechanism \cite{87Zr_shear}, chiral doublet bands \cite{chirality1, chirality2}, and, more recently, multiple chiral doublet bands \cite{chiraldoublet} associated with octupole correlations. These characteristics make this mass region one of the richest laboratories for exploring the interplay between collective and single particle degrees of freedom. 
Of particular interest are nuclei in the vicinity of Z=38–40 and N=50, where the proton subshell structure and the robust N=50 neutron shell closure strongly influence the evolution of nuclear states. For Yttrium isotopes in this vicinity, the competition between inert core behavior and particle–hole excitations across these shell boundaries plays a crucial role in shaping level schemes and generating high angular momenta states, similar to A$\sim$132 mass region~\cite{132sn1, 132sn2}. Consequently, these nuclei provide a stringent testing ground for shell-model interactions and model spaces developed for the A$\sim$80 mass region. Detailed investigations of their spectroscopic properties therefore offer valuable benchmarks for refining our microscopic understanding of nuclear structure and the underlying nuclear forces.

The experimental investigations of the high spin structure of $^{89}$Y was carried out using in-beam $\gamma$-ray spectroscopy \cite{89Y_expt1,89Y_expt2,89Y_expt3}. A significant study was reported by Funke \textit{et al}. \cite{funke}, in which the level scheme was initially extended up to $31/2^{+}$ through the $^{87}$Rb($\alpha$, 2n) reaction. Positive parity states associated with neutron core excitations were identified in this work. Based on $\gamma$–$\gamma$ coincidence analysis and lifetime measurements, transition probabilities for several states were extracted. Subsequently, Li \etal ~\cite{Li2016}, using the heavy-ion fusion reaction ~$^{82}Se (^{11}B,4n)$ at beam energies of 48 and 52 MeV, significantly extended the level scheme up to an excitation energy of approximately 11.5 MeV. In particular, negative parity states associated with neutron core excitations around 7 MeV were identified for the first time in \(^{89}\)Y. These experimental developments provide an extensive spectroscopic database for investigating the evolution of the high spin structure and the role of cross shell neutron excitations in this nucleus.

Several theoretical investigations have been performed to understand the spectroscopic properties of this nucleus. In particular, the nuclear SM has been extensively employed to characterize its underlying nuclear structure. Previous SM studies of $^{89}$Y have primarily been performed within the $fpg_{9/2}$ model space  \cite{funke,pc}. In the pioneering work of Reif \etal  ~\cite{funkeII}, an enlarged model space comprising the proton $1f_{5/2}$, $2p_{3/2}$, $2p_{1/2}$, and $1g_{9/2}$ orbitals together with the neutron $2p_{1/2}$, $1g_{9/2}$, $2d_{5/2}$, and $1g_{7/2}$ orbitals was employed. Both excitation energies and electromagnetic transition probabilities were investigated in that work. Subsequently, Srivastava \etal ~\cite{pc} presented a SM description restricted to the low spin states. More recently, Li \etal~\cite{Li2016} employed the cross shell interaction to investigate both the low- and high spin structure of $^{89}$Y. Das \etal~\cite{IEMPHYS2026} subsequently investigated the low-lying level structure and seniority properties of $^{89}$Y using different effective interaction Hamiltonians. Despite these efforts, a systematic comparative study of the sensitivity of the calculated nuclear structure to different effective interactions, neutron cross-shell-model spaces, and excitation truncations remains desirable. In particular, the interplay between proton excitations and neutron-core excitations, as reflected in the wave-function compositions, orbital occupation probabilities, and electromagnetic transition strengths across a broad spin range, warrants further investigation.

Motivated by these observations, in the present work, we perform a systematic SM investigation of \(^{89}\mathrm{Y}\) using different effective interactions and model-space extensions to examine the role of neutron excitations across the \(N=50\) shell closure. 
In contrast to considering neutron-core excitation as a
single structural mechanism, we explore different truncation schemes
involving neutron promotions to the \(1g_{7/2}\) and \(2d_{5/2}\) orbitals,
as well as higher-order particle-hole excitations. This enables us to
investigate the orbital dependence of cross-shell excitations and assess
their influence on the evolution of nuclear structure with angular momentum. In addition to the excitation spectra, we analyze the wave-function compositions, orbital occupation probabilities, and proton-neutron angular-momentum couplings to identify the microscopic mechanisms responsible for the formation of the observed states. 
The calculated \(\mathrm{B(E2)}\) and \(\mathrm{B(M1)}\) transition strengths are also examined to provide complementary insights into the structural changes associated with neutron-core excitation. Through this combined analysis, we aim to establish the relative importance of proton excitation and neutron-core excitation and identify the model space restrictions that provide the most consistent description of the structure of \(^{89}\mathrm{Y}\).

This paper is organized as follows. The model space, effective interactions, and computational procedure are described in Sec. \ref{effective interaction}. Section \ref{Results} presents a detailed comparison of the calculated level scheme with the available experimental data along with the deviation between these two. A microscopic analysis of the individual states, including wave function analysis (Subsec. \ref{wavefunction}) and the occupation probabilities of the relevant shells (Subsec. \ref{occupation}), is also presented. Furthermore, the electromagnetic transition probabilities between different levels of $^{89}$Y are examined in terms of their underlying nuclear structure (Subsec. \ref{Transition}). These analyses provide insights into the role of neutron core excitations in the evolution of the high spin structure. Finally, a summary of the present work is given in Sec. \ref{Summary}.
\section{\label{effective interaction} Effective interactions}

In this section, we describe the effective interactions employed to investigate the high spin states of $^{89}$Y. SM calculations have been performed using the \textsc{NuShellX@MSU} code~\cite{nushellx} with four different effective interactions, namely \ju ~\cite{jun45}, \gle ~\cite{glekpn}, \sn ~\cite{snt}, and \gwb ~\cite{gwbxg1, gwbxg2, gwbxg3, gwbxg4}. These interactions employ different model spaces and therefore provide complementary descriptions of the structure of $^{89}$Y, particularly with respect to neutron excitations across the $N=50$ shell closure.

The \ju ~interaction \cite{jun45} is widely used to describe the nuclear structure in the mass region around $A\sim 90$ \cite{jun45_ref1}. It is a realistic interaction derived from the Bonn-C potential and is defined in the $f_{5/2}pg_{9/2}$ model space (jj44pn) with $^{56}\mathrm{Ni}$ as an inert core. The active orbitals,
$1f_{5/2}$, $2p_{3/2}$, $2p_{1/2}$, and $1g_{9/2}$, are considered for both protons and neutrons, with single particle energies (SPEs) of $-8.7087$, $-9.8280$, $-7.8388$, and $-6.2617$ MeV, respectively. The interaction contains 133 two-body matrix elements (TBMEs) and four SPEs, which were determined empirically by fitting to experimental data in the mass region $A=63$--$96$. For $^{89}\mathrm{Y}$, the neutron $f_{5/2}pg_{9/2}$ valence space is completely occupied at the $N=50$ shell closure. Consequently, neutron excitations within and across the $N=50$ gap cannot be described within the restricted jj44pn model space. The calculation with \ju ~is therefore performed in the full $f_{5/2}pg_{9/2}$ model space without any additional truncation.

To incorporate neutron excitations across the $N=50$ shell closure, three cross shell interactions with extended model spaces are also employed. The \gle ~interaction \cite{glekpn} spans the $\pi fpg_{9/2}$ and $\nu gds$ model spaces with $^{60}\mathrm{Ca}$ as an inert core. The proton model space consists of the $1f_{7/2}$, $1f_{5/2}$, $2p_{1/2}$, $2p_{3/2}$, and $1g_{9/2}$ orbitals, with corresponding SPEs of $-8.085$, $-10.840$, $-11.361$, $-9.610$, and $-5.424$ MeV, respectively. The neutron model space includes the $1g_{9/2}$, $1g_{7/2}$, $2d_{5/2}$, $2d_{3/2}$, and $3s_{1/2}$ orbitals, with SPEs of $-10.164$, $10.927$, $4.220$, $7.217$, and $4.371$ MeV, respectively. In this present work, the proton $1f_{7/2}$ orbital is assumed to be completely occupied. Thus, the active proton space effectively becomes similar to that of the jun45pn interaction, while neutron excitations from the $\nu1g_{9/2}$ orbital across the $N=50$ shell gap are permitted.

The \sn ~interaction \cite{snt} provides an even more extended model space (sn) in which neutron excitations across the $N=50$ shell closure can originate from the $\nu f_{5/2}pg_{9/2}$ orbitals. The model space consists of the $1f_{5/2}$, $2p_{3/2}$, $2p_{1/2}$, $1g_{9/2}$, $1g_{7/2}$, $2d_{5/2}$, $2d_{3/2}$, and $3s_{1/2}$ orbitals for both protons and neutrons, with $^{56}\mathrm{Ni}$ as the inert core. The proton SPEs for these orbitals are $0.525$, $1.228$, $5.106$, $5.518$, $20.656$, $18.893$, $20.016$, and $16.895$ MeV, respectively. For the neutron orbitals below the $N=50$ shell closure, the SPEs are taken as zero, while those above the shell closure, namely $1g_{7/2}$, $2d_{5/2}$, $2d_{3/2}$, and $3s_{1/2}$, have SPEs of $4.352$, $2.313$, $3.440$, and $2.772$ MeV, respectively.

The \gwb ~cross shell effective interaction is also employed in the present study. The \gwb ~effective interaction is constructed with different interactions \cite{gwbxg1, gwbxg2, gwbxg3, gwbxg4}. The original set of 974 two-body matrix elements (TBMEs) was obtained from the bare \(G\)-matrix derived from the H7B potential \cite{gwbxg1}. This interaction considers a proton model space similar to that of \ju, together with an extended neutron model space containing the $2p_{1/2}$ and $1g_{9/2}$ orbitals below the $N=50$ shell gap and the $1g_{7/2}$, $2d_{5/2}$, $2d_{3/2}$, and $3s_{1/2}$ orbitals above the shell gap with $^{68}Ni$ core. The single particle energies for proton 1$f_{5/2}$, 2$p_{3/2}$, 2$p_{1/2}$, 1$g_{9/2}$ are -5.322, -6.144, -3.941, -1.250 MeV respectively. For neutron SPES for 2$p_{1/2}$, 1$g_{9/2}$, 1$g_{7/2}$, 2$d_{5/2}$, 2$d_{3/2}$, and 3$s_{1/2}$  are 
 -0.696, -2.597, 5.159, 1.830, 4.261 and 1.741 Mev respectively. The inclusion of these cross shell configurations substantially increases the dimensionality of the shell-model basis. Consequently, unrestricted calculations are computationally prohibitive, making it essential to impose physically motivated truncation schemes.

For all three cross shell interactions, proton excitations across the $N=50$ shell gap are not allowed. The calculations are performed using four neutron particle-hole truncation schemes, which are denoted as follows: restriction I -  0p0h: no neutron is excited across the $N=50$ shell gap; restriction II - 1p1h ($1g_{7/2}$): one neutron is excited across the $N=50$ shell gap into the $1g_{7/2}$ orbital; restriction III - 1p1h ($2d_{5/2}$): one neutron is excited across the $N=50$ shell gap into the $2d_{5/2}$ orbital; restriction IV - 2p2h ($1g_{7/2}$--$2d_{5/2}$): at most two neutrons are promoted across the $N=50$ shell gap, with a maximum occupancy of one neutron in each of the $1g_{7/2}$ and $2d_{5/2}$ orbitals. These truncation schemes allow us to systematically examine the effects of neutron cross shell excitations on the calculated high spin states of $^{89}$Y while keeping the SM dimensions computationally tractable.

\section{\label{Results} Results and Discussion}

\subsection{Excitation energies and RMS deviation}

\begin{figure}
\centering
\includegraphics[width=1.0\linewidth]{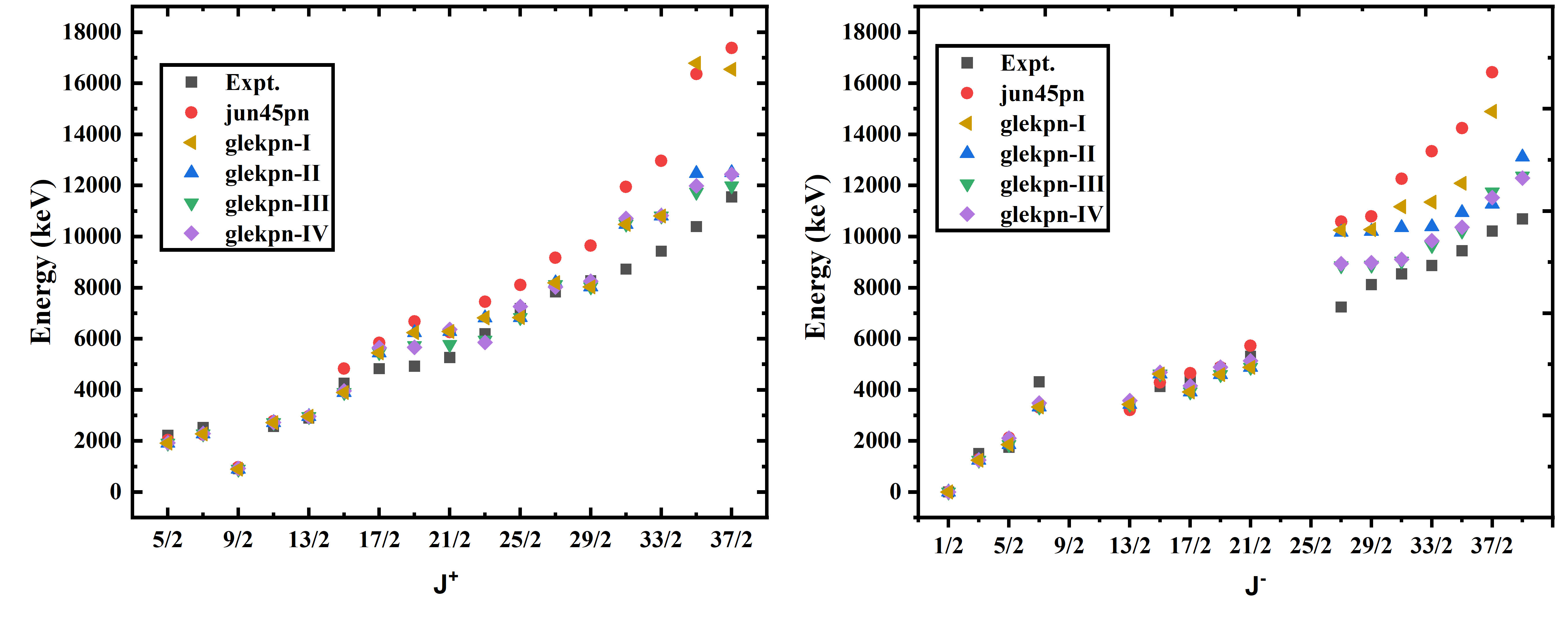}
\caption{(Color Online) Comparison of experimentally obtained  positive parity (left) and negative parity (right) yrast states of $^{89}$Y with theoretical calculation with \gle ~effective interaction in different truncation schemes. For details of the legends, see text.}
\label{glekpn_all_restrictions}
\end{figure}
\begin{figure}
\centering
\includegraphics[width=1.0\linewidth]{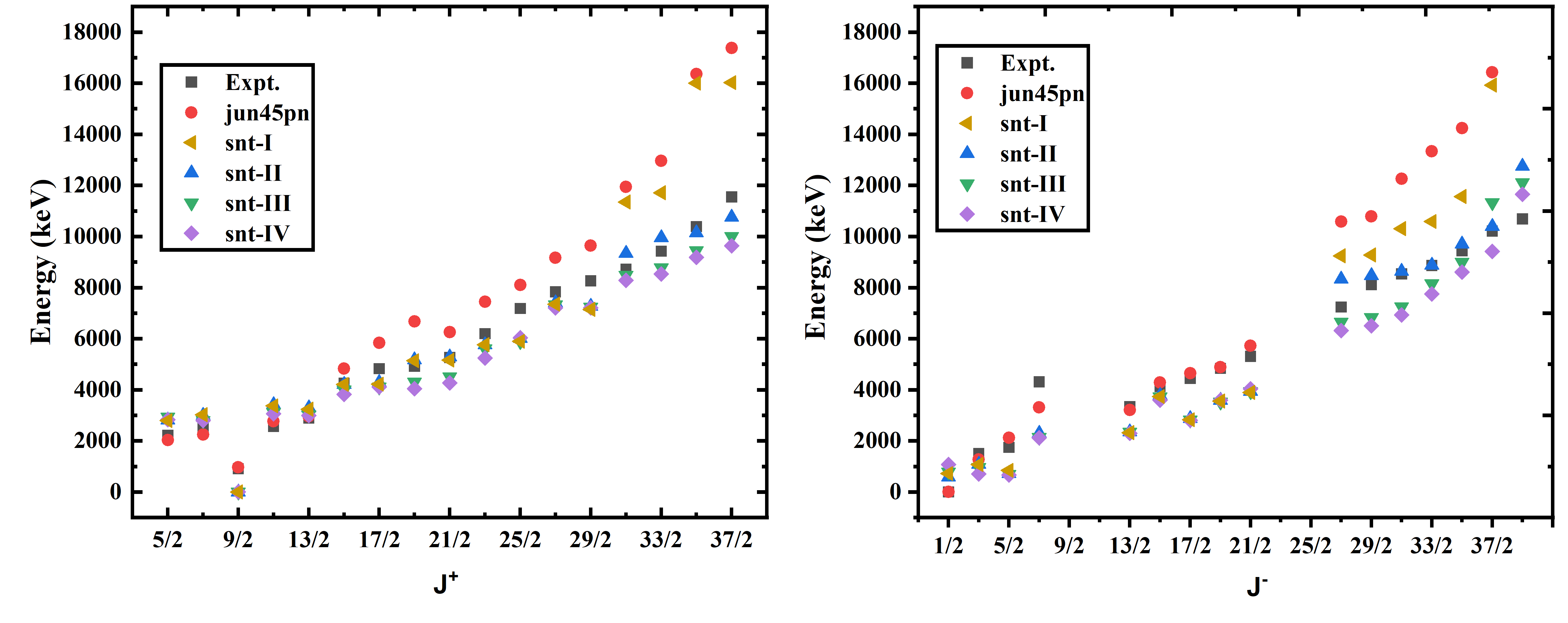}
\caption{Same as Fig. \ref{glekpn_all_restrictions} with \sn \, interaction. }
\label{snt_all_restrictions}
\end{figure}
\begin{figure}
\centering
\includegraphics[width=1.0\linewidth]{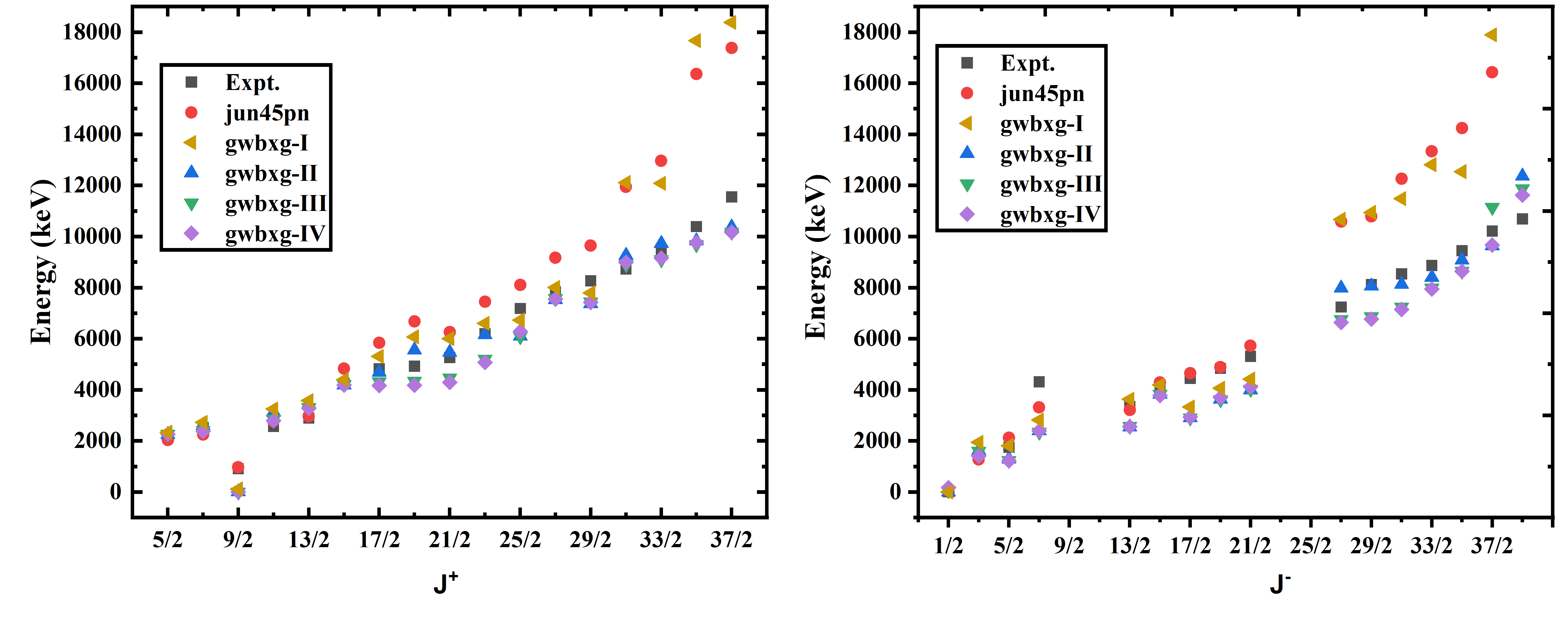}
\caption{Yrast states with \gwb ~interaction. Same as Fig. \ref{glekpn_all_restrictions}.}
\label{gwbxg_all_restrictions}
\end{figure}

The excitation energies of the yrast positive and negative parity states of $^{89}$Y were calculated using the \ju, \gle, \sn, and \gwb ~interactions. Appropriate truncation schemes were imposed on the \gle, \sn, and \gwb ~interactions to investigate the influence of neutron excitations across the $N=50$ shell closure. The calculated spectra are compared with the experimental data in Figs.~\ref{glekpn_all_restrictions}, \ref{snt_all_restrictions}, and \ref{gwbxg_all_restrictions}. All the states are yrast, unless mentioned.

\subsubsection{Positive Parity States}

\noindent{Low-spin region ($5/2^+ - 15/2^+$):} The low spin positive parity states are reproduced reasonably well by all interactions and truncation schemes. The \ju ~interaction provides a satisfactory description of the spectrum up to approximately the $13/2^+$ state, indicating that the dominant configurations of these levels are largely confined within the restricted $f_{5/2}pg_{9/2}$ model space. A similar behaviour is obtained with the \gle-I, corresponding to the $0p0h$ configuration, where neutron excitations across the $N=50$ shell gap are not permitted. The resulting spectrum remains close to that obtained with \ju, with deviations typically within $\sim 300$ keV, and reproduces the experimental levels reasonably well up to about $15/2^+$.

The inclusion of the $\nu1g_{7/2}$ orbital through the \gle-II truncation leads to only minor modifications in this spin region. Owing to the relatively large single particle energy of this orbital, its occupancy remains small, and the calculated excitation energies closely follow those obtained with \gle-I. Similarly, the \gle-III and \gle-IV calculations introduce only marginal changes, typically of the order of a few tens of keV, indicating that cross shell excitations play a limited role in the structure of these low spin states.

A different trend is observed for the \sn~Hamiltonian. In the $0p0h$ configuration (\sn-I), the $5/2^+$ and $7/2^+$ states are overpredicted by approximately 500 keV. More importantly, this truncation fails to reproduce the experimental ground state and instead predicts the $9/2^+$ state as the ground state, underestimating its excitation energy by nearly 900 keV. Conversely, the $11/2^+$ state is overpredicted by a comparable amount. The agreement improves for the $13/2^+$ state, while the $15/2^+$ level is reproduced remarkably well, with deviations of only a few tens of keV. The inclusion of particle hole excitations in the \sn-II, \sn-III, and \sn-IV calculations does not alter the predicted ground state configuration. However, for the excited states, these truncation schemes systematically lower the calculated excitation energies by approximately 300 keV, resulting in improved agreement with experiment for levels up to $13/2^+$. As a consequence of this overall downward shift, the $15/2^+$ state becomes slightly underpredicted in these calculations.

The \gwb ~interaction yields a better overall description of the low spin spectrum than the \sn ~Hamiltonian. The \gwb-I calculation correctly predicts the $9/2^+$ state as the first excited state, consistent with the experimental observation, although its excitation energy is underestimated. Excellent agreement is obtained for the $5/2^+$, $7/2^+$, and $15/2^+$ states. In contrast, the $11/2^+$ and $13/2^+$ levels are overpredicted within the \gwb-I scheme. The remaining truncation schemes further reduce the excitation energy of the $9/2^+$ state, resulting in its prediction as the ground state. Noticeable differences are observed for the $7/2^+$ and $11/2^+$ states, where the inclusion of particle hole excitations lowers the excitation energies by approximately 200 keV. However, it provides a satisfactory description of the remaining low spin positive parity states and predict excitation energies that are generally similar to those obtained with \gwb-I. 

Overall, the agreement obtained with all three interactions suggests that the low spin positive parity states are predominantly governed by configurations involving orbitals below the $N=50$ shell closure, while cross shell excitations contribute only weakly to their structure.\\

\noindent{Intermediate spin region ($17/2^+$--$25/2^+$):} Differences among the interactions become increasingly evident in the intermediate spin region. The \ju ~interaction continues to overpredict the overall evolution of the spectrum, and the discrepancy with experiment gradually increases with spin, reaching up to 1.3 MeV. A similar trend is observed for the \gle-I calculation, where the agreement deteriorates beyond approximately $15/2^+$ and persists up to the $23/2^+$ state. However, the deviations remain smaller than those obtained with the \ju ~interaction. The $25/2^+$ state is reproduced with better agreement, with a deviation of only $\sim$250 keV.

Allowing one neutron excitation into the $\nu1g_{7/2}$ orbital through the \gle-II truncation results in a modest improvement. However, the most pronounced effect is obtained with the \gle-III calculation, where neutron excitations into the lower lying $\nu2d_{5/2}$ orbital become possible. The calculated excitation energies follow the experimental trend more closely and reproduce the $23/2^+$ and $25/2^+$ states with very good agreement, with deviations of only 200--300 keV. This demonstrates the increasing importance of neutron excitations across the shell gap. The \gle-IV calculation yields results very similar to those obtained with \gle-III for the $23/2^+$ and $25/2^+$ states, indicating that the dominant structural effects are already captured within the $1p1h$ excitation scheme.

For the \sn ~interaction, the \sn-I calculation predicts excitation energies very close to the experimental values but underpredicts the $25/2^+$ state by almost 1 MeV. The inclusion of $1p1h$ excitations in the \sn-II calculation provides almost the same results as the \sn-I truncation. The \sn-III and \sn-IV calculations produce a slightly more compressed spectrum, lowering the excitation energies by approximately 600 keV.

The \gwb-I calculation exhibits behaviour different from that of \sn-I, overpredicting the excitation energies by 500--1000 keV. The \gwb-II calculation lowers the excitation energies and reproduces the experimental trend more closely. The \gwb-III calculation further lowers the excitation energies, which in turn leads to an underprediction of the experimental levels, while the \gwb-IV truncation introduces only minor additional changes.\\

\noindent{High spin region ($27/2^+$ - $37/2^+$):} The high spin positive parity states provide the most sensitive test of neutron excitations across the $N=50$ shell closure. In this region, the limitations of the restricted model spaces become apparent. The \ju ~interaction increasingly overestimates the excitation energies, and the discrepancy reaches several ($\sim$ 6 MeV) MeV for the highest spin states. On the other side, \gle-I produces the  $27/2^+$ and $29/2^+$ states in very good agreement. However, it starts to over predict the above high spin states and deviation reaches up to ~4 MeV. The \gle-II truncation follows the \gle-I up to $31/2^+$ state but improves the description of the spectrum for the higher states, reducing the discrepancy for the highest spin states to approximately 1 MeV. A substantially better agreement is obtained with the \gle-III calculation and reproduce the observed trend more accurately. The \gle-IV calculation remains nearly identical to \gle-III, demonstrating that the essential features of the high spin positive parity structure are already accounted for by low order cross shell excitations.

The \sn-I calculation under predicts the $27/2^+$ and $29/2^+$ states, whereas it overestimates the high spin excitation energies. The inclusion of cross shell excitations in \sn-II produces a clear improvement, while \sn-III and \sn-IV yield only moderate additional compression of the spectrum. A similar evolution is observed for the \gwb ~interaction. The \gwb-I calculation exhibits large deviations from experiment, whereas \gwb-II and \gwb-III provide a much improved description of the high spin states. The \gwb-IV results remain close to those obtained with \gwb-III, indicating that higher order particle hole excitations contribute only modest refinements to the calculated excitation energies.

\subsubsection{Negative Parity States}

\noindent{Low spin region ($1/2^-$ - $21/2^-$):} The low spin negative parity states are generally reproduced satisfactorily by all interactions. The \ju ~interaction provides a good description of the experimentally observed levels up to approximately $21/2^-$. Similar agreement is obtained with the \gle-I calculation, confirming that the dominant configurations of these states are largely confined below the $N=50$ shell closure. The inclusion of the $\nu1g_{7/2}$ orbital in the \gle-II calculation produces only moderate changes, while the \gle-III and \gle-IV truncations have little influence on the calculated energies in this spin range. All of the truncations of the \sn ~interactions show almost similar spectrum and under predicts the experimental level energies in this low spin region. The \gwb ~interactions exhibit comparable behaviour. 

\noindent{Intermediate-spin region ($J^\pi \geq 23/2^-$):} As the angular momentum increases, the influence of neutron excitations above the $N=50$ shell closure becomes increasingly apparent. The \ju ~interaction begins to overestimate the excitation energies, and the discrepancy grows steadily up to 6 MeV with spin.

The \gle-I calculation shows a similar tendency, while the \gle-II truncation provides an improvement for the higher spin states. A significantly better agreement is obtained with the \gle-III calculation, and reproduce the experimental trend more closely. The \gle-IV results remain nearly identical to those of \gle-III.

For the \sn ~interaction, the inclusion of cross shell excitations lowers the excitation energies and produces a more compressed spectrum. Although the highest spin states tend to lie slightly below the experimental values with truncation \sn-III and  \sn-IV, the agreement remains significantly better than that obtained without cross shell excitations.

A similar trend is observed for the \gwb ~interaction. The progressive improvement from \gwb-I to \gwb-IV demonstrates the importance of neutron particle hole excitations across the $N=50$ shell closure in establishing the structure of the high spin negative parity states. Among the truncation schemes considered, \gwb-IV provides the closest overall agreement with the experimental negative parity yrast sequence.

Overall, the results obtained with the \gle, \sn, and \gwb ~interactions consistently indicate that the low spin states can be described primarily within configurations below the $N=50$ shell closure, whereas a realistic description of the high spin positive and negative parity yrast sequences requires the inclusion of neutron excitations into orbitals above the shell gap.


To quantify the overall agreement between the calculated and experimental excitation energies obtained with the different interaction Hamiltonians and truncation schemes, the root mean square (rms) deviation was evaluated using experimentally established states (yrast and non-yrast) with confirmed spin and parity assignments \cite{ensdf} according to the given formula

\begin{equation} \label{rms}
\mathrm{R.M.S. \ deviation} = \sqrt{\frac{1}{n} \sum_{i=1}^{n} \left( E^{i}_{\mathrm{expt.}} - E^{i}_{\mathrm{theo.}} \right)^2}.
\end{equation}

Here, E$^{i}_{\mathrm{theo.}}$ and E$^{i}_{\mathrm{expt.}}$ denote the calculated and experimental excitation energies, respectively, n is the number of yrast states considered for the calculation.

\begin{figure}
\vspace{-2cm}

\includegraphics[width =1\linewidth]{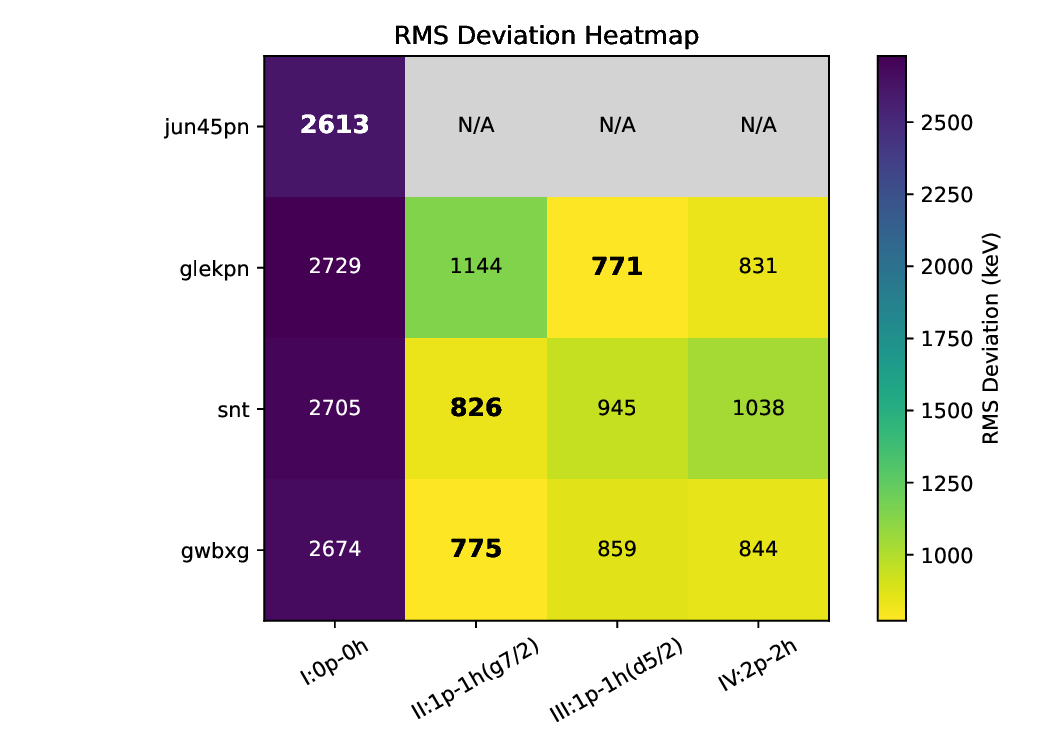}

\caption{\label{rmsfig} Heatmap representation of R.M.S. deviations (in keV) for different effective interactions and truncation schemes. Lighter shades correspond to lower rms values, indicating better agreement with experimental data, while darker shades represent larger deviations. Bold entries denote the optimal truncation for each interaction. Grey cells indicate cases where calculations are not available.}
\end{figure}

Figure \ref{rmsfig} shows that the $0p0h$ calculations, i.e., those without cross shell excitations, yield the largest rms deviations for all interactions. The inclusion of neutron excitations across the $N=50$ shell gap leads to a systematic reduction in the rms deviation for all cross shell interactions. The minimum rms values are obtained for the \gle-III truncation, \sn-II truncation and \gwb-II truncation. Among all interactions and truncation schemes, the lowest rms deviation is obtained for the \gle-III calculation. Since cross shell excitations are not possible within the \ju ~interaction, the corresponding entries are indicated by grey cells.

\begin{figure}
\centering
\includegraphics[width=1.0\linewidth]{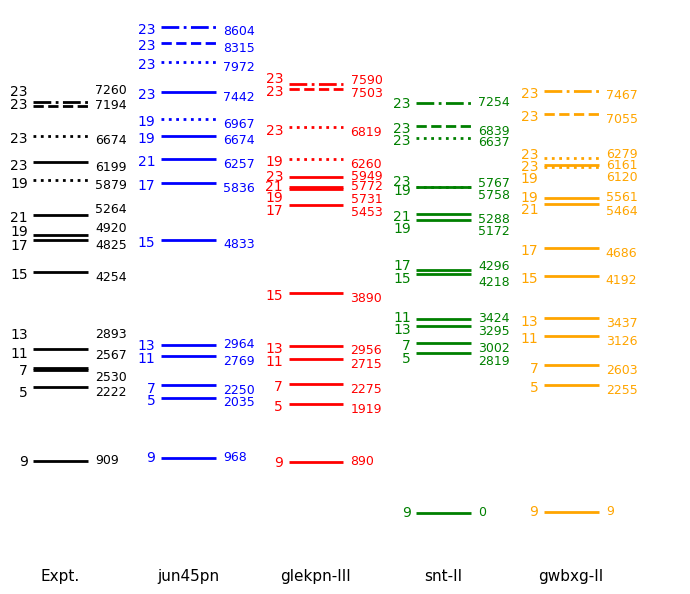}
\caption{Comparison of SM obtained level scheme, calculated with different effective interactions with the experimentally observed levels for positive parity states up to 2J$^{\pi}$ = 23$^+$. The mentioned spin-parity are in 2J$^{\pi}$ format.}
\label{pos_parity}
\end{figure}

\begin{figure}
\centering
\includegraphics[width=1.0\linewidth]{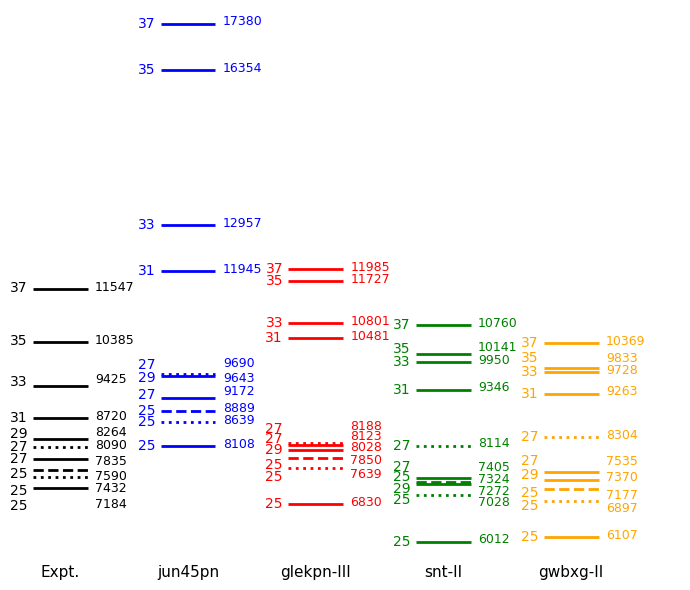}
\caption{Same as Fig. \ref{pos_parity}, for positive parity states from 2J$^{\pi}$ = 25$^+$ to 2J = 37$^+$.}
\label{pos_parity_ext}
\end{figure}

\begin{figure}
\centering
\includegraphics[width=1.0\linewidth]{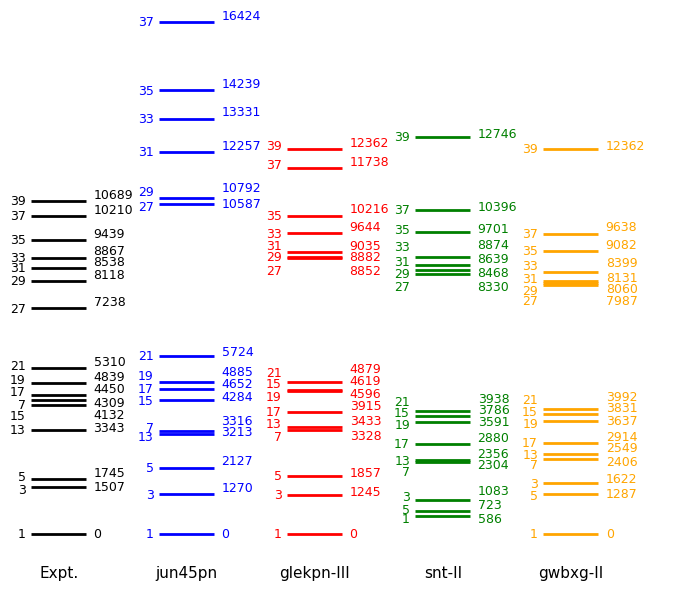}
\caption{Same as Fig. \ref{pos_parity}, for negative parity states up to 2J$^{\pi}$ = 39$^-$.}
\label{neg_parity}
\end{figure}

Based on the rms analysis, the interaction and truncation schemes providing the best overall agreement were selected for a more detailed comparison with the experimental level scheme. Figures \ref{pos_parity} and \ref{pos_parity_ext} compare the calculated positive parity levels with the experimental data, while Fig. \ref{neg_parity} presents the corresponding comparison for the negative parity states obtained using the \ju, \gle-III, \sn-II, and \gwb-II interactions. The comparison of the excitation energies of positive parity states have been shown in different figures for better clarity.

Experimentally, the ground state of $^{89}$Y is assigned a spin-parity of $1/2^{-}$, while the first excited state corresponds to the well known $9/2^{+}$ spin trap isomer. This state decays to the ground state through an $M4+E5$ transition, emitting a $\sim$909 keV $\gamma$ ray. Both the \ju~interaction and all truncation schemes of the \gle~interaction reproduce the correct ordering and excitation energies of these two states. The \sn-II interaction, however, fails to preserve the observed sequence. In this case, the $9/2^{+}$ state becomes the ground state, while the $1/2^{-}$ level is shifted upward to approximately 0.5 MeV and appears as the first excited state. The \gwb-II truncation scheme retains the correct ordering of the states, although the excitation energy of the $9/2^{+}$ level is considerably underestimated.

The \ju ~interaction provides a reasonable description of the low spin spectrum up to the $13/2^{+}$ state. Beyond this region, the calculated excitation energies gradually deviate from the experimental values, leading to an increasing overestimation of the higher spin levels. The discrepancy reaches approximately 1.4 MeV for the non yrast $23/2^{+}_{3}$ state and increases to nearly 6 MeV for the $37/2^{+}$ state. For the \gle-III interaction, the $5/2^{+}$, $7/2^{+}$, $11/2^{+}$, and $13/2^{+}$ states are reproduced well, while the $15/2^{+}$ level is predicted slightly below the experimental value. The $17/2^{+}$, $19/2^{+}$, $21/2^{+}$, and non yrast $19/2^{+}$ states are obtained at somewhat higher energies, with deviations of approximately 0.4--0.7 MeV, although their ordering is preserved. Good agreement is obtained for both the yrast $23/2^{+}$ and non yrast $23/2^{+}_{2}$ states. The non yrast $23/2^{+}_{3}$ and $23^{+}_{4}$ levels are placed slightly above the experimental values. The yrast $25/2^{+}$, $27/2^{+}$, and $29/2^{+}$ states are marginally underestimated, whereas the $31/2^{+}$, $33/2^{+}$, $35/2^{+}$, and $37/2^{+}$ levels are overestimated by about 0.5 MeV. The non yrast $25/2^{+}_{2}$ and $27/2^{+}_{2}$ states are also reproduced with good accuracy. In the case of the \sn-II interaction, the yrast $5/2^{+}$, $7/2^{+}$, $11/2^{+}$, and $13/2^{+}$ levels are predicted approximately 0.6 MeV above the experimental values, whereas the $15/2^{+}$ and $17/2^{+}$ states are underestimated. The yrast and non yrast states between $19/2^{+}$ and $23/2^{+}$ are reproduced reasonably well in terms of excitation energy. Nevertheless, several changes in the level ordering are observed, particularly for the $11/2^{+}$--$13/2^{+}$ and $15/2^{+}$--$17/2^{+}$ pairs. Both the yrast and non yrast $25/2^{+}$ states are significantly underestimated. Furthermore, the calculated higher spin levels from $31/2^{+}$ to $37/2^{+}$ are compressed into an energy interval that is nearly half of the experimentally observed span. The \gwb-II interaction exhibits behaviour broadly similar to that of the \sn-II interaction. Although the overall energy span is somewhat expanded and the agreement with experiment improves in several cases, a noticeable compression of the high spin spectrum remains. This effect becomes particularly evident for the $25/2^{+}$ to $37/2^{+}$ states, including the corresponding non yrast levels. As a result, the excitation energies of these high spin states are systematically underestimated.

\subsection{\label{wavefunction} Wavefunction Analysis}
As noted from the energy comparison, proton excitations alone provide a reasonable description of the excitation energies only up to $\sim$4 MeV. Therefore, in this section, the wave functions corresponding to the $0p$–$0h$ truncation of the cross shell interactions are not analyzed, except for \ju ~interaction (Table \ref{wavefn_jun_pos} and \ref{wavefn_jun_neg}). The tables  represent the wavefunction of restrictions I, II and III of \gle ~(Table \ref{wavefn_glekpn_pos} and Table \ref{wavefn_glekpn_neg}), \sn ~(Table \ref{wavefn_snt_pos} and Table \ref{wavefn_snt_neg}), \gwb ~(Table \ref{wavefn_gwbxg_pos} and Table \ref{wavefn_gwbxg_neg}) interaction,  respectively.

The ground state $1/2^{-}$ is predominantly characterized by the $\pi 2p_{1/2}^{1}$ configuration with seniority $\nu=1$ across all interactions. The first excited $9/2^{+}$ isomer is also a seniority 1 state, mainly arising from the single particle $\pi 1g_{9/2}^{1}$ excitation. The subsequent positive parity states with $5/2^{+}_1 \leq J^\pi \leq 11/2^{+}_1$ (excluding $9/2^{+}_{1}$) are dominated by the $\pi 2p_{3/2}^{-1}2p_{1/2}^{1}1g_{9/2}^{1}$ configuration in all interactions. The $13/2^{+}$ yrast state is largely generated from this configuration in \ju ~and all truncations of \gle \, and \gwb ~(III, IV) interactions. This observation matches with SM predicted in Reif \etal However, different configurations, namely $\pi 1f_{5/2}^{-2}1g_{9/2}^{3}$ for \sn\, (both $1p$–$1h$ and $2p$–$2h$ excitations) and  for $\pi 1f_{5/2}^{-1}2p_{1/2}^{1}1g_{9/2}^{1}$ for \gwb-II respectively, are predicted. Experimentally, an energy gap of $\sim$1.4 MeV is observed between the $13/2^{+}_{1}$ and $15/2^{+}_{1}$ state. The $15/2^{+}_1$ state is best reproduced by the \sn-II interaction and is predominantly generated from the proton configuration $\pi 1f_{5/2}^{-1}2p_{3/2}^{-1}1g_{9/2}^{3}$. In contrast, for \ju,~\gle\,(II–IV), and \gwb~II, III interactions favor another proton driven configuration $\pi 1f_{5/2}^{-1}2p_{1/2}^{1}1g_{9/2}^{1}$. However, in the $2p$–$2h$ truncation of the \gwb ~interaction, this state exhibits mainly neutron core excitation with the leading configuration $\pi 1g_{9/2}^{1} \otimes \nu 1g_{9/2}^{-1}2d_{5/2}^{1}$. In general, all these states are seniority 3 state. \\
The subsequent $17/2^{+}_1$ state is over predicted by $\sim$1 MeV in the \ju ~interaction, with $\pi 1f_{5/2}^{-1}2p_{3/2}^{-1}1g_{9/2}^{1}$ as a most contributing configuration. In the \gle ~interaction, the $1p$–$1h$ truncation indicates a predominantly proton driven structure, while the $2p$–$2h$ truncation shows a transition to neutron core excitation with  $\pi 1g_{9/2}^{1} \otimes \nu 1g_{9/2}^{-1}2d_{5/2}^{1}$ partition, which is consistent with Reif \etal ~\cite{funkeII}. Both $1p-1h$ and $2p-2h$ excitation of \sn ~and $1p1h-1g_{7/2}$ of \gwb ~calculations favor proton dominated configurations, mainly $\pi 1f_{5/2}^{-2}1g_{9/2}^{3}$. However, the $1p1h-1g_{7/2}$ and $2p2h$ of \gwb ~show behavior similar to that of $2p-2h$ excitation of \gle, indicating an increasing role of neutron excitations.

The next yrast $19/2^{+}$ state, experimentally observed at 4920.46 keV, is over-predicted by $\sim$1.7 MeV in the \ju ~interaction. This rises predominantly from the proton configuration $\pi 2p_{3/2}^{-2}1g_{9/2}^{3}$ with a $\sim$68 \% amplitude. For the cross shell interactions, the $1p$–$1h$ ($2d_{5/2}$) and $2p$–$2h$ truncations indicate a neutron core excited structure coupled with proton excitations. In contrast, the $1p$–$1h$ ($1g_{7/2}$) truncation predicts a mainly proton driven configuration, with $\pi 1f_{5/2}^{-1}2p_{3/2}^{-1}1g_{9/2}^{1}$ as a leading component. This behavior reflects the relatively large energy gap between the $\nu 1g_{7/2}$ and $\nu 1g_{9/2}$ orbitals, making the $\nu 1g_{9/2}^{-1}1g_{7/2}^{1}$ excitation energetically unfavorable leading to proton driven excitation. A similar behavior is observed for the $21/2^{+}_{1}$ and $23/2^{+}_{1}$ states also. The states formed through neutron core excitation are mainly seniority 3 states. However, those are generated through proton excitation mixture of seniority 3 and 5. In the higher energy region ($\sim$7–8 MeV), the states $25/2^{+}_1$, $27/2^{+}_1$ and $29/2^{+}_1$ are predominantly proton driven ($\pi1f_{5/2}^{-2}1g_{9/2}^{3}$) across most of truncation schemes of the cross shell interactions with increasing seniority number. This suggests that, in this energy range, the energy required for proton pair breaking and neutron core excitation becomes comparable. 

The next high spin states, $31/2^{+} \leq I^\pi \leq 37/2^{+}$ state, are generally generated through this proton configuration coupled with neutron core excitation, with dominant structures $\pi 1f_{5/2}^{-2}1g_{9/2}^{3} \otimes \nu 1g_{9/2}^{-1}2d_{5/2}^{1}$ for the $1p$–$1h$ ($2d_{5/2}$) and $2p$–$2h$ truncations, and $\pi 1f_{5/2}^{-2}1g_{9/2}^{3} \otimes \nu 1g_{9/2}^{-1}1g_{7/2}^{1}$ for the $1p$–$1h$ ($1g_{7/2}$) truncation. An exception is observed in the \gle ~interaction for the $31/2^{+}$ and $33/2^{+}$ states. The $31/2^{+}$ state mainly arises from $\pi 1f_{5/2}^{-3}2p_{1/2}^{1}1g_{9/2}^{3}$ in \gle-II, III and from $\pi 1f_{5/2}^{-1}2p_{1/2}^{1}1g_{9/2}^{1} \otimes \nu 1g_{9/2}^{-1}1g_{7/2}^{1}$ in \gle-IV, whereas the $33/2^{+}$ state remains proton driven with the configuration $\pi 1f_{5/2}^{-2}2p_{3/2}^{-1}2p_{1/2}^{1}1g_{9/2}^{3}$ across all truncations.
The $37/2^{+}_1$ state represents the highest positive parity level observed experimentally. Following the $\Delta J = 1$ sequence, the next expected state is the $39/2^{+}_1$ level, which is predominantly generated from the configuration $\pi 1f_{5/2}^{-2}1g_{9/2}^{3} \otimes \nu 1g_{9/2}^{-1}2d_{5/2}^{1}$ for the $1p$–$1h$ ($2d_{5/2}$) and $2p$–$2h$ truncations, and from $\pi 1f_{5/2}^{-2}1g_{9/2}^{3} \otimes \nu 1g_{9/2}^{-1}1g_{7/2}^{1}$ in the $1p$–$1h$ ($1g_{7/2}$) truncation. These states are mostly seniority 5 or 7 states.

The low lying yrast negative parity states, such as $5/2^{-}$ and $7/2^{-}$, are predominantly described by the configurations $\pi 1f_{5/2}^{-1}$ and $\pi 1f_{5/2}^{-1}1g_{9/2}^{2}$ (or $\pi 1p_{3/2}^{-1}1g_{9/2}^{2}$), respectively, across all interactions. The $13/2^{-}$ and $15/2^{-}$ states mainly arise from the $\pi 1p_{3/2}^{-1}1g_{9/2}^{2}$ and $\pi 1f_{5/2}^{-1}1g_{9/2}^{2}$ configurations. The higher-spin states with $17/2^{-}_1$ to $21/2^{-}_1$ exhibit a common dominant structure $\pi 1f_{5/2}^{-1}1g_{9/2}^{2}$.
Up to this state, the yrast states are predominantly generated through proton excitations, and the calculated excitation energies show reasonable agreement with the experimental data across all interactions. 

The next higher-spin state, $27/2^{-}$, is experimentally observed at an excitation energy of $\sim$ 7 MeV. As noted earlier, a better agreement with experiment is obtained for the cross shell interactions compared to \ju ~interaction. It reflects the increasing importance of neutron excitations in this energy region. The $27/2^{-}$ and $29/2^{-}$ states are predominantly characterized by the proton configurations $\pi 1p_{3/2}^{-1}1g_{9/2}^{2}$ and $\pi 1f_{5/2}^{-1}1g_{9/2}^{2}$, coupled with $\nu 1g_{9/2}^{-1}2d_{5/2}^{1}$ for the $1p$–$1h$ ($2d_{5/2}$) and $2p$–$2h$ truncations, and with $\nu 1g_{9/2}^{-1}1g_{7/2}^{1}$ for the $1p$–$1h$ ($1g_{7/2}$) truncation across all cross shell interactions. Within the \ju ~interaction, the states from $27/2^{-}$ to $33/2^{-}$ are primarily characterized by proton excitations involving the promotion of two protons from the $\pi 2p_{3/2}$ orbital to the $\pi 1g_{9/2}$ orbital, with the dominant configuration $\pi 1f_{5/2}^{-1}2p_{3/2}^{-2}1g_{9/2}^{4}$. Similarly, the states with $31/2^{-} \leq I^\pi \leq 37/2^{-}$ are mainly generated from the $\pi 1f_{5/2}^{-1}1g_{9/2}^{2}$ proton configuration coupled to the same neutron partitions in cross shell interactions. An exception is observed for the $37/2^{-}$ state in the $1p$–$1h$ ($2d_{5/2}$) truncation, where the proton configuration $\pi 1f_{5/2}^{-1}1p_{3/2}^{-1}1p_{1/2}^{1}1g_{9/2}^{2}$ becomes dominant. The highest negative parity state, $39/2^{-}$, is also described by similar proton configurations coupled with neutron core excitation. The decomposition of proton and neutron angular momenta further indicates that both proton and neutron excitations contribute comparably to the generation of these high spin states, highlighting the essential role of cross shell neutron excitations beyond the $N=50$ shell closure. $39/2^-$ state can not be generated through proton excitation only in $fpg_{9/2}$ model space.


\begin{table*}[!t]
\centering

\begin{minipage}[t]{0.49\textwidth}
\centering
\caption{Total angular momenta ($2J^+$) and their corresponding proton ($2J_p$)  angular-momentum decompositions and their probability (Prob.) (considering only components with probabilities greater than 10\%) is noted. For each decomposition, the main proton configurations of the wavefunction and their contribution (Cont.) and seniority number (Sen.) of the particular configuration for positive parity yrast states of $^{89}Y$ using \ju ~interaction are noted. $\pi$ WF represents the proton occupancy in ($1f_{5/2}$, $2p_{3/2}$, $2p_{1/2}$, $1g_{9/2}$) orbitals, respectively.}
\label{wavefn_jun_pos}
\vspace{0.3cm}

\resizebox{\textwidth}{!}{
\begin{tabular}{ccccccc}
\hline
 & \multicolumn{4}{c}{Config. of $\pi$ WF} &
 Total & Seniority \\
\cmidrule(r){2-5}
$\mathbf{2J^+}$ & $p_1$ & $p_2$ & $p_3$ & $p_4$ &
 Con. (\%) & Number \\
\hline
\hline
$\mathbf{9}$  & 6 & 4 & 0 & 1 & 60.05 & 1 \\
   & 5 & 4 & 1 & 1 & 12.80 & 3 \\
   & 6 & 2 & 0 & 3 & 12.44 & 1 \\
$\mathbf{11}$ & 6 & 3 & 1 & 1 & 58.75 & 3 \\
   & 5 & 4 & 1 & 1 & 15.11 & 3 \\
$\mathbf{13}$ & 6 & 3 & 1 & 1 & 63.77 & 3 \\
$\mathbf{15}$ & 5 & 4 & 1 & 1 & 77.10 & 3 \\
   & 5 & 2 & 1 & 3 & 12.11 & 3 \\
$\mathbf{17}$ & 5 & 3 & 2 & 1 & 41.10 & 3 \\
   & 4 & 4 & 2 & 1 & 19.80 & 3 \\
   & 6 & 2 & 0 & 3 & 16.80 & 3 \\
$\mathbf{19}$ & 6 & 2 & 0 & 3 & 68.80 & 3 \\
   & 6 & 1 & 1 & 3 & 14.14 & 3 \\
$\mathbf{21}$ & 6 & 2 & 0 & 3 & 71.62 & 3 \\
$\mathbf{23}$ & 5 & 3 & 0 & 3 & 46.70 & 5 \\
   & 6 & 2 & 0 & 3 & 20.53 & 5 \\
$\mathbf{25}$ & 6 & 2 & 0 & 3 & 34.13 & 5 \\
   & 5 & 3 & 0 & 3 & 29.47 & 5 \\
$\mathbf{27}$ & 5 & 3 & 0 & 3 & 77.87 & 5 \\
$\mathbf{29}$ & 5 & 3 & 0 & 3 & 77.80 & 5 \\
$\mathbf{31}$ & 5 & 2 & 1 & 3 & 88.88 & 7 \\
$\mathbf{33}$ & 4 & 3 & 1 & 3 & 92.61 & 7 \\
$\mathbf{35}$ & 4 & 2 & 0 & 5 & 90.22 & 7 \\
$\mathbf{37}$ & 4 & 2 & 0 & 5 & 93.94 & 9 \\
   & 3 & 3 & 0 & 5 & 4.79 & 9 \\
\hline
\hline
\end{tabular}
}
\end{minipage}
\hfill
\begin{minipage}[t]{0.49\textwidth}
\centering
\caption{The details of the yrast negative parity states obtained with the \ju ~interaction are presented below. The columns follow the same format and conventions as those in Table \ref{wavefn_jun_pos}.}
\label{wavefn_jun_neg}
\vspace{0.3cm}

\resizebox{\textwidth}{!}{
\begin{tabular}{ccccccc}
\hline
 & \multicolumn{4}{c}{Config. of $\pi$ WF} &
 Total & Seniority \\
\cmidrule(r){2-5}
$\mathbf{2J^-}$ & $p_1$ & $p_2$ & $p_3$ & $p_4$ &
 Con. (\%) & Number \\
\hline
\hline
$\mathbf{1}$  & 6 & 4 & 1 & 0 & 71.52 & 1 \\
   & 6 & 2 & 1 & 2 & 13.66 & 1 \\
$\mathbf{13}$ & 6 & 3 & 0 & 2 & 77.55 & 3 \\
$\mathbf{15}$ & 6 & 3 & 0 & 2 & 73.30 & 3 \\
$\mathbf{17}$ & 6 & 3 & 0 & 2 & 67.85 & 3 \\
$\mathbf{19}$ & 6 & 3 & 0 & 2 & 52.99 & 3 \\
   & 5 & 4 & 0 & 2 & 24.89 & 3 \\
$\mathbf{21}$ & 5 & 4 & 0 & 2 & 66.05 & 3 \\
   & 5 & 2 & 2 & 2 & 10.44 & 3 \\
$\mathbf{27}$ & 5 & 2 & 0 & 4 & 65.36 & 5 \\
   & 6 & 1 & 0 & 4 & 14.00 & 5 \\
$\mathbf{29}$ & 5 & 2 & 0 & 4 & 78.96 & 5 \\
$\mathbf{31}$ & 5 & 2 & 0 & 4 & 76.30 & 7 \\
$\mathbf{33}$ & 5 & 2 & 0 & 4 & 65.49 & 7 \\
   & 4 & 3 & 0 & 4 & 12.98 & 7 \\
$\mathbf{35}$ & 4 & 3 & 0 & 4 & 90.38 & 7 \\
$\mathbf{37}$ & 4 & 2 & 1 & 4 & 90.36 & 9 \\
\hline
\hline
\end{tabular}
}
\end{minipage}

\end{table*}

\begin{sidewaystable}
\centering 
\caption{Total angular momenta ($2J^+$) and their corresponding proton ($2J_p$) and neutron ($2J_n$) angular-momentum decompositions and their probability (Prob.)( considering only components with probabilities greater than 10\%) is noted. For each decomposition, the main proton  and neutron  configurations of the wavefunction
and their contribution (Cont.) and seniority number (Sen.) of the particular configuration for positive parity yrast
states of $^{89}Y$ using \gle ~interaction for three different restrictions are noted. $\pi$ WF represents the proton occupancy in
($1f_{5/2}$, $2p_{3/2}$, $2p_{1/2}$, $1g_{9/2}$) orbitals, respectively. $\nu$ WF gives  the neutron occupancy
in  ($1g_{9/2}$, $1g_{7/2}$, $2d_{5/2}$) orbitals, respectively. Restriction details given in text.
}
\vspace{0.5cm}
\setlength{\tabcolsep}{1pt}
\resizebox{0.8\textwidth}{!}{
\begin{tabular}{c|ccccccc|ccccccc|ccccccc}
\hline
\multicolumn{22}{c}{}\\
\multicolumn{8}{c|}{\gle-II(1p1h-1g$_{7/2}$)} & \multicolumn{7}{c|}{\gle-III(1p1h-2d$_{5/2}$)} & \multicolumn{7}{c}{\gle-IV(2p2h)}\\
\multicolumn{22}{c}{}\\
\hline
\multicolumn{22}{c}{}\\
$\mathbf{2J^+}$ & $2J_p$ & $2J_n$ & Prob.(\%) & $\pi$ WF & $\nu$ WF & Cont.(\%) & Sen. & $2J_p$ & $2J_n$ & Prob.(\%) & $\pi$ WF & $\nu$ WF & Cont.(\%) & Sen. & $2J_p$ & $2J_n$ & Prob.(\%) & $\pi$ WF & $\nu$ WF & Cont.(\%) & Sen.\\
\multicolumn{22}{c}{}\\
\hline
\hline
$\mathbf{9}$ & 9 & 0 & 100 & (6, 4, 0, 1) & (10, 0, 0) & 58.30 & 1 & 9 & 0 & 100 & (6, 4, 0, 1) & (10, 0, 0) & 58.29 & 1 &   9 & 0 & 100 & (6, 4, 0, 1) & (10, 0, 0) & 59.34 & 1   \\

&  &  &  & (6, 2, 2, 1) & (10, 0, 0) & 14.19 & 1 &  &  &  & (6, 2, 2, 1) & (10, 0, 0) & 14.19 & 1 &    &  &  & (6, 2, 2, 1) & (10, 0, 0) & 14.38 & 1   \\
 
&  &  &  & (4, 4, 2, 1) & (10, 0, 0) & 11.38 & 1 &  &  &  & (4, 4, 2, 1) & (10, 0, 0) & 11.36 & 1 &  &    &  & (4, 4, 2, 1) & (10, 0, 0) & 10.90 & 1   \\
\hline
$\mathbf{11}$ & 11 & 0 & 100 & (6, 3, 1, 1) & (10, 0, 0) & 79.50 & 3 & 11 & 0 & 100  & (6, 3, 1, 1) & (10, 0, 0) & 79.50 & 3  & 11 & 0 & 100 & (6, 3, 1, 1) & (10, 0, 0) & 80.45 & 3   \\
\hline
$\mathbf{13}$ & 13 & 0 & 100 & (6, 3, 1, 1) & (10, 0, 0) & 68.86 & 3 & 13 & 0 & 100 & (6, 3, 1, 1) & (10, 0, 0) & 68.85 & 3 &   13 & 0 & 100 & (6, 3, 1, 1) & (10, 0, 0) & 70.62 & 3   \\

&  &  &  & (6, 2, 2, 1) & (10, 0, 0) & 10.59 & 3 &  &  &  & (6, 2, 2, 1) & (10, 0, 0) & 10.53 &  &  &  &    & (6, 2, 2, 1) & (10, 0, 0) & 10.77 & 3   \\
\hline
$\mathbf{15}$ & 15 & 0 & 100 & (5, 4, 1, 1) & (10, 0, 0) & 67.28 & 3 & 15 & 0 & 100 & (5, 4, 1, 1) & (10, 0, 0) & 67.35 & 3 &  15 & 0 & 100 & (5, 4, 1, 1) & (10, 0, 0) & 69.57 & 3  \\

&  &  &  & (5, 3, 2, 1) & (10, 0, 0) & 13.57 & 3 &  &  &  & (5, 3, 2, 1) & (10, 0, 0) & 13.57 & 3 &  &  &  &   (5, 3, 2, 1) & (10, 0, 0) & 13.69 & 3   \\
 
&  &  &  & (5, 2, 1, 3) & (10, 0, 0) & 10.83 & 3 &  &  &  & (5, 2, 1, 3) & (10, 0, 0) & 10.81 & 3 &  &  &  &   (5, 2, 1, 3) & (10, 0, 0) & 10.01 & 3  \\
\hline
$\mathbf{17}$ & 17 & 0 & 100 & (4, 4, 0, 3) & (10, 0, 0) & 58.02 & 3 & 17 & 0 & 99.86 & (4, 4, 0, 3) & (10, 0, 0) & 58.60  & 3 &  9 & 10 & 42.26 & (6, 4, 0, 1) & (9, 0, 1) & 25.20 & 3   \\

&  &  &  & (5, 3, 0, 3) & (10, 0, 0) & 11.79 & 3 &  &  &  & (5,3,0,3) & (10, 0, 0) & 11.31   & 3 & 9 & 12 & 32.63 & (6, 4, 0, 1) & (9, 0, 1) & 19.36 & 3   \\
 
&  &  &  &  &  &  &  &  &  &  &  &  &    &  & 9 & 8 & 17.45 & (6, 4, 0, 1) & (9, 0, 1) & 10.43 & 3   \\
\hline
$\mathbf{19}$ & 19 & 0 & 100 & (5, 3, 0, 3) & (10, 0, 0) & 42.60 & 5 & 9 & 12 & 57.68 & (6, 4, 0, 1) & (9, 0, 1) & 34.24 & 3 &  9 & 12 & 58.70 & (6, 4, 0, 1) & (9, 0, 1) & 35.05 & 3  \\

&  &  &  & (4, 4, 0, 3) & (10, 0, 0) & 27.98 & 3 & 9 & 10 & 29.42 & (6, 4, 0, 1) & (9, 0, 1) & 17.40 & 3 &  9 & 10 & 28.68 & (6, 4, 0, 1) & (9, 0, 1) & 17.04 & 3   \\
 
&  &  &  & (4, 3, 1, 3) & (10, 0, 0) & 10.48 & 5 &  &  &  &    &  &  &  &  &  &  &  &  &  &    \\
\hline
$\mathbf{21}$ & 21 & 0 & 100 & (4, 4, 0, 3) & (10, 0, 0) & 68.75 & 3 & 9 & 12 & 73.03 & (6, 4, 0, 1) & (9, 0, 1) & 43.38 & 3 &   9 & 14 & 64.06 & (6, 4, 0, 1) & (9, 0, 1) & 37.83 & 3   \\

&  &  &  & (4, 3, 1, 3) & (10, 0, 0) & 12.82 & 5 &  &  &  & (6, 2, 2, 1) & (9, 0, 1) & 10.24 & 3 &  9 & 12 & 24.65 & (6, 4, 0, 1) & (9, 0, 1) & 13.97 & 3   \\
 
&  &  &  &  &  &  &  & 9 & 14 & 22.71 & (6, 4, 0, 1) & (9, 0, 1) & 13.29 & 3 &   &  &  &  &  &  &    \\
\hline
$\mathbf{23}$ & 23 & 0 & 100 & (5, 3, 0, 3) & (10, 0, 0) & 48.23 & 5 & 9 & 14 & 95.24 & (6, 4, 0, 1) & (9, 0, 1) & 56.34 & 3 &  9 & 14 & 95.06 & (6, 4, 0, 1) & (9, 0, 1) & 56.39 & 3   \\

 &  &  &  & (4, 4, 0, 3) & (10, 0, 0) & 26.06 & 5 &  &  &  & (6, 2, 2, 1) & (9, 0, 1) & 13.23 & 3 &  &    &  & (6, 2, 2, 1) & (9, 0, 1) & 13.14 & 3   \\
 
 &  &  &  & (4, 3, 1, 3) & (10, 0, 0) & 10.35 & 5 &  &  &  & (4, 4, 2, 1) & (9, 0, 1) & 11.82 & 3 &  &  &   & (4, 4, 2, 1) & (9, 0, 1) & 11.25 & 3   \\
\hline
$\mathbf{25}$ & 25 & 0 & 100 & (4, 4, 0, 3) & (10, 0, 0) & 63.4 & 5 & 25 & 0 & 99.99 & (4, 4, 0, 3) & (10, 0, 0) & 63.63 & 5 &   25 & 0 & 99.97 & (4, 4, 0, 3) & (10, 0, 0) & 69.15 & 5  \\

&  &  &  & (4, 3, 1, 3) & (10, 0, 0) & 20.35 & 5 &  &  &  & (4, 3, 1, 3) & (10, 0, 0) & 20.29 & 5 &  &    &  & (4, 3, 1, 3) & (10, 0, 0) & 21.91 & 5   \\
\hline

\hline
\end{tabular}}
\label{wavefn_glekpn_pos}
\end{sidewaystable}

\begin{sidewaystable}
\centering
\caption*{contd.}
\vspace{0.5cm}
\setlength{\tabcolsep}{1pt}
\resizebox{0.8\textwidth}{!}{
\begin{tabular}{c|ccccccc|ccccccc|ccccccc}
\hline
\multicolumn{22}{c}{}\\
\multicolumn{8}{c|}{\gle-II(1p1h-1g$_{7/2}$)} & \multicolumn{7}{c|}{\gle-III(1p1h-2d$_{5/2}$)} & \multicolumn{7}{c}{\gle-IV(2p2h)}\\
\multicolumn{22}{c}{}\\
\hline
\multicolumn{22}{c}{}\\
$\mathbf{2J^+}$ & $2J_p$ & $2J_n$ & Prob.(\%) & $\pi$ WF & $\nu$ WF & Cont.(\%) & Sen. & $2J_p$ & $2J_n$ & Prob.(\%) & $\pi$ WF & $\nu$ WF & Cont.(\%) & Sen. & $2J_p$ & $2J_n$ & Prob.(\%) & $\pi$ WF & $\nu$ WF & Cont.(\%) & Sen.\\
\multicolumn{22}{c}{}\\
\hline
\hline

$\mathbf{27}$ & 27 & 0 & 100 & (4, 4, 0, 3) & (10, 0, 0) & 48.03 & 5 & 13 & 14 & 99.01 & (6, 3, 1, 1) & (9, 0, 1) & 76.41 & 5 &   13 & 14 & 99.10 & (6, 3, 1, 1) & (9, 0, 1) & 75.67 & 5   \\

&  &  &  & (4, 3, 1, 3) & (10, 0, 0) & 21.26 & 7 &  &  &    &  &  &  &  &  &  &  &  &  &  &    \\
\hline
$\mathbf{29}$ & 29 & 0 & 100 & (4, 4, 0, 3) & (10, 0, 0) & 61.83 & 5 & 29 & 0 & 100 & (4, 4, 0, 3) & (10, 0, 0) & 61.92 & 5 &  29 & 0 & 99.99 & (4, 4, 0, 3) & (10, 0, 0) & 65.34 & 5   \\

&  &  &  & (4, 3, 1, 3) & (10, 0, 0) & 25.74 & 7 &  &  &  & (4, 3, 1, 3) & (10, 0, 0) & 25.69 & 7 &  &   &  & (4, 3, 1, 3) & (10, 0, 0) & 25.62 & 7   \\
\hline
$\mathbf{31}$ & 31 & 0 & 100 & (3, 4, 1, 3) & (10, 0, 0) & 70.66 & 7 & 31 & 0 & 99.95 & (3, 4, 1, 3) & (10, 0, 0) & 70.02 & 7 &  15 & 16 & 98.86 & (5, 4, 1, 1) & (9, 1, 0) & 67.74 & 5   \\

&  &  &  & (3, 3, 2, 3) & (10, 0, 0) & 13.66 & 7 &  &  &  & (3, 3, 2, 3) & (10, 0, 0) & 13.47 & 7 &  &  &   & (5, 2, 1, 3) & (9, 1, 0) & 11.75 & 5   \\
 
&  &  &  & (4, 3, 1, 3) & (10, 0, 0) & 11.31 &  7&  &  &  & (4, 3, 1, 3) & (10, 0, 0) & 11.97 & 7 &  &  &    & (5, 3, 2, 1) & (9, 1, 0) & 11.04 & 5   \\

\hline
$\mathbf{33}$ & 33 & 0 & 100 & (4, 3, 1, 3) & (10, 0, 0) & 79.15 & 7 & 33 & 0 & 99.99 & (4, 3, 1, 3) & (10, 0, 0) & 79.16 & 7 &  33 & 0 & 99.99 & (4, 3, 1, 3) & (10, 0, 0) & 79.14 & 7   \\

&  &  &  & (4, 2, 2, 3) & (10, 0, 0) & 18.7 & 7 &  &  &  & (4, 2, 2, 3) & (10, 0, 0) & 18.68 & 7 &  &  &    & (4, 2, 2, 3) & (10, 0, 0) & 18.79 & 7  \\
\hline
$\mathbf{35}$ & 21 & 14 & 64.47 & (4, 4, 0, 3) & (9, 1, 0) & 44.19 & 5 & 21 & 14 & 92.33 & (4, 4, 0, 3) & (9, 0, 1) & 62.19 & 5 &  21 & 14 & 90.21 & (4, 4, 0, 3) & (9, 0, 1) & 62.53 & 7   \\

&  &  &  & (4, 3, 1, 3) & (9, 1, 0) & 11.05 & 7 &  &  &  &  &  &  &  &  &  &   & (4, 3, 1, 3) & (9, 0, 1) & 11.18 & 7   \\
 
& 21 & 16 & 26.66 & (4, 4, 0, 3) & (9, 1, 0) & 18.20 & 5 &  &  &  &  &  &    &  &  &  &  &  &  &  &    \\

\hline
$\mathbf{37}$ & 21 & 16 & 93.7 & (4, 4, 0, 3) & (9, 1, 0) & 64.27 & 5 & 25 & 12 & 67.08 & (4, 4, 0, 3) & (9, 0, 1) & 42.19 & 7 &  25 & 12 & 71.15 & (4, 4, 0, 3) & (9, 0, 1) & 48.99 & 7   \\

&  &  &  & (4, 3, 1, 3) & (10, 0, 0) & 16.50 & 7 &  25 & 14 & 30.66 & (4, 4, 0, 3) & (9, 0, 1) & 18.69 & 7 &  25 & 14 & 24.8 & (4, 4, 0, 3) & (9, 0, 1) & 16.77 & 7   \\
\hline
$\mathbf{39}$ & 25 & 14 & 57.83 & (4, 4, 0, 3) & (9, 1, 0) & 33.87 & 5 & 25 & 14 & 97.66 & (4, 4, 0, 3) & (9, 0, 1) & 60.71 & 7 & 25 & 14 & 96.75 & (4, 4, 0, 3) & (9, 0, 1) & 66.57 & 7  \\

&  &  &  & (4, 3, 1, 3) & (9, 1, 0) & 14.04 & 7 &  &  &  & (4, 3, 1, 3) & (9, 0, 1) & 19.67 & 7 &  &    &  & (4, 3, 1, 3) & (9, 0, 1) & 20.73 & 7   \\

& 25 & 16 & 40.74 & (4, 4, 0, 3) & (9, 1, 0) & 23.87 & 5 &  &  &    &  &  &  &  &  &  &  &  &  &  &    \\
\hline
\hline
\end{tabular}}
\end{sidewaystable}

\begin{sidewaystable}
\centering
\caption{The details of the yrast negative parity states obtained with the \gle ~interaction are presented below. The columns follow the same format and conventions as those in Table \ref{wavefn_glekpn_pos}.}
\vspace{0.5cm}
\setlength{\tabcolsep}{1pt}
\resizebox{0.8\textwidth}{!}{
\begin{tabular}{c|ccccccc|ccccccc|ccccccc}
\hline
\multicolumn{22}{c}{}\\
\multicolumn{8}{c|}{\gle-II(1p1h-1g$_{7/2}$)} & \multicolumn{7}{c|}{\gle-III(1p1h-2d$_{5/2}$)} & \multicolumn{7}{c}{\gle-IV(2p2h)}\\
\multicolumn{22}{c}{}\\
\hline
\multicolumn{22}{c}{}\\
$\mathbf{2J^-}$ & $2J_p$ & $2J_n$ & Prob.(\%) & $\pi$ WF & $\nu$ WF & Cont.(\%) & Sen. & $2J_p$ & $2J_n$ & Prob.(\%) & $\pi$ WF & $\nu$ WF & Cont.(\%) & Sen. & $2J_p$ & $2J_n$ & Prob.(\%) & $\pi$ WF & $\nu$ WF & Cont.(\%) & Sen.\\
\multicolumn{22}{c}{}\\
\hline
\hline
$\mathbf{1}$ & 1 & 0 & 100 & (6,4,1,0) & (10,0,0) & 80.50 & 1 & 1 & 0 & 100 & (6,4,1,0) & (10,0,0) & 80.49 & 1 & 1 & 0 & 100 & (6,4,1,0) & (10,0,0) & 80.49 & 1 \\
 &  &  &  & (6,2,1,2) & (10,0,0) & 14.87 & 1 &  &  &  & (6,2,1,2) & (10,0,0) & 14.87 & 1 &  &  &  & (6,2,1,2) & (10,0,0) & 14.84 & 1 \\
\hline
$\mathbf{13}$& 13 & 0 & 100 & (6,3,0,2) & (10,0,0) & 48.05 & 3 & 13 & 0 & 100 & (6,3,0,2) & (10,0,0) & 48.03 & 3 & 13 & 0 & 100 & (6,3,0,2) & (10,0,0) & 51.68 & 3 \\
 &  &  &  & (5,4,0,2) & (10,0,0) & 27.67 & 3 &  &  &  & (5,4,0,2) & (10,0,0) & 27.69 & 3 &  &  &  & (5,4,0,2) & (10,0,0) & 25.49 & 3 \\
\hline
$\mathbf{15}$ & 15 & 0 & 100 & (5,4,0,2) & (10,0,0) & 32.05 & 3 & 15 & 0 & 99.99 & (5,4,0,2) & (10,0,0) & 31.88 & 3 & 15 & 0 & 100 & (6,3,0,2) & (10,0,0) & 30.95 & 3 \\
 &  &  &  & (6,3,0,2) & (10,0,0) & 23.35 & 3 &  &  &  & (6,3,0,2) & (10,0,0) & 23.51 & 3 &  &  &  & (5,4,0,2) & (10,0,0) & 21.26 & 3 \\
 &  &  &  & (5,3,1,2) & (10,0,0) & 14.26 & 3 &  &  &  & (5,3,1,2) & (10,0,0) & 14.22 & 5 &  &  &  & (5,3,1,2) & (10,0,0) & 15.27 & 5 \\
 &  &  &  & (4,4,1,2) & (10,0,0) & 11.08 & 3 &  &  &  & (4,4,1,2) & (10,0,0) & 11.11 & 3 &  &  &  & (4,4,1,2) & (10,0,0) & 11.65 & 3 \\
\hline
$\mathbf{17}$ & 17 & 0 & 100 & (5,4,0,2) & (10,0,0) & 68.13 & 3 & 17 & 0 & 100 & (5,4,0,2) & (10,0,0) & 68.14 & 3 & 17 & 0 & 100 & (5,4,0,2) & (10,0,0) & 62.74 & 3 \\
 &  &  &  &  &  &  &  &  &  &  &  &  &  &  &  &  &  & (5,3,1,2) & (10,0,0) & 11.97 & 5 \\
\hline
$\mathbf{19}$ & 19 & 0 & 100 & (5,4,0,2) & (10,0,0) & 70.36 & 3 & 19 & 0 & 100 & (5,4,0,2) & (10,0,0) & 70.35 & 3 & 19 & 0 & 100 & (5,4,0,2) & (10,0,0) & 71.67 & 3 \\
 &  &  &  &  &  &  &  &  &  &  &  &  &  &  &  &  &  & (5,3,1,2) & (10,0,0) & 12.87 & 5 \\
\hline
$\mathbf{21}$ & 21 & 0 & 100 & (5,4,0,2) & (10,0,0) & 65.77 & 3 & 21 & 0 & 100 & (5,4,0,2) & (10,0,0) & 65.77 & 3 & 21 & 0 & 100 & (5,4,0,2) & (10,0,0) & 61.73 & 3 \\
 &  &  &  &  &  &  &  &  &  &  &  &  &  &  &  &  &  & (4,4,1,2) & (10,0,0) & 15.68 & 5 \\
\hline
$\mathbf{27}$ & 13 & 14 & 65.13 & (6,3,0,2) & (9,1,0) & 28.63 & 5 & 13 & 14 & 57.05 & (6,3,0,2) & (9,0,1) & 22.16 & 5 & 17 & 12 & 45.59 & (5,4,0,2) & (9,0,1) & 25.98 & 5 \\
 &  &  &  & (5,4,0,2) & (9,1,0) & 17.72 & 5 &  &  &  & (5,4,0,2) & (9,0,1) & 18.95 & 5 &  17 & 10 & 24.37 & (5,4,0,2) & (9,0,1) & 14.03 & 5  \\
 & 13 & 16 & 28.45 & (6,3,0,2) & (9,1,0) & 12.67 & 5 & 17 & 10 & 15.91 & (5,4,0,2) & (9,0,1) & 10.24 & 5 & 17 & 14 & 13.37 & (5,4,0,2) & (9,0,1) & 7.20 & 5\\
 &  &  & & &  &  & & 17 & 12 & 12.11 & (5,4,0,2) & (9,0,1) & 7.32 & 5 &  &  &  &  &  &  &  \\
 \hline
 
$\mathbf{29}$ & 13 & 16 & 89.04 & (6,3,0,2) & (9,1,0) & 38.85 & 5 & 17 & 12 & 66.40 & (5,4,0,2) & (9,0,1) & 43.12 & 5 & 17 & 12 & 65.05 & (5,4,0,2) & (9,0,1) & 38.80 & 5 \\
 &  &  &  & (5,4,0,2) & (9,1,0) & 24.23 & 5 & 17 & 14 & 25.23 & (5,4,0,2) & (9,0,1) & 15.45 & 5 & 17 & 14 & 26.39 & (5,4,0,2) & (9,0,1) & 14.42 & 5 \\
\hline
$\mathbf{31}$ & 17 & 14 & 60.92 & (5,4,0,2) & (9,1,0) & 34.79 & 5 & 17 & 14 & 88.52 & (6,3,0,2) & (9,0,1) & 57.54 & 5 & 17 & 14 & 88.93 & (5,4,0,2) & (9,0,1) & 51.19 & 5 \\
 & 17 & 16 & 33.88 & (5,4,0,2) & (9,1,0) & 19.07 & 5 &  &  &  &  &  &  &  &  &  &  & (5,3,1,2) & (9,0,1) & 10.13 & 5 \\
 
\hline
$\mathbf{33}$& 17 & 16 & 94.09 & (5,4,0,2) & (9,1,0) & 54.14 & 5 & 19 & 14 & 81.58 & (5,4,0,2) & (9,0,1) & 55.17 & 5 & 19 & 14 & 76.53 & (5,4,0,2) & (9,0,1) & 54.53  & 5 \\
 &  &  &  & (5,3,1,2) & (9,1,0) & 12.68 & 5 & 21 & 12 & 13.15 & (5,4,0,2) & (9,0,1) & 8.17 & 5 & 21 & 12 & 15.97 & (5,4,0,2) & (9,0,1) & 9.39 & 5 \\

\hline
$\mathbf{35}$ & 19 & 16 & 94.2 & (5,4,0,2) & (9,1,0) & 60.81 & 5 & 21 & 14 & 97.13 & (5,4,0,2) & (9,0,1) & 64.25 & 5 & 21 & 14 & 95.40 & (5,4,0,2) & (9,0,1) & 58.25 &5  \\
 &  &  &  & (5,3,1,2) & (9,1,0) & 11.6 & 5 &  &  &  &  &  &  &  &  &  &  & (4,4,1,2) & (9,0,1) &  17.32& 7 \\

 \hline
$\mathbf{37}$ & 21 & 16 & 98.58 & (5,4,0,2) & (9,1,0) & 62.17 & 5 & 23 & 14 & 70.34 & (5,3,1,2) & (9,0,1) & 53.09 & 7 & 21 & 16 & 95.26 & (5,4,0,2) & (9,1,0) & 56.04 & 5 \\
 &  &  &  & (5,3,1,2) & (9,1,0) & 12.68 & 5 & 25 & 12 & 23.07 & (5,3,1,2) & (9,0,1) & 15.79 & 7 &  &  &  & (4,4,1,2) & (9,1,0) & 15.79 & 7 \\
 &  &  &  &  &  &  &  &  &  &  &  &  &  &  &  &  &  & (5,3,1,2) & (9,1,0) & 12.31 & 7 \\

\hline
$\mathbf{39}$ & 23 & 16 & 88.78 & (5,3,1,2) & (9,1,0) & 55.27 & 7 & 25 & 14 & 99.71 & (5,3,1,2) & (9,0,1) & 66.5 & 7 & 25 & 14 & 98.46 & (5,3,1,2) & (9,0,1) & 63.48  & 7 \\
 &  &  &  & (4,4,1,2) & (9,1,0) & 17.11 & 7 &  &  &  & (4,4,1,2) & (9,0,1) & 14.92 & 7 &  &  &  & (4,4,1,2) & (9,0,1) & 16.19 & 7  \\
 &  &  &  &  &  &  &  &  &  &  & (5,2,2,2) & (9,0,1) & 12.41 & 7 &  &  &  & (5,2,2,2) & (9,0,1) & 16.19 & 7 \\

\hline
\hline
\end{tabular}}
\label{wavefn_glekpn_neg}
\end{sidewaystable}

\begin{sidewaystable}
\centering
\caption{The details of the yrast positive parity states obtained with the \sn ~interaction are presented below. The columns follow the same format and conventions as those in Table \ref{wavefn_glekpn_pos}.}
\vspace{0.5cm}
\setlength{\tabcolsep}{1pt}
\resizebox{0.8\textwidth}{!}{%
\begin{tabular}{c|ccccccc|ccccccc|ccccccc}
\hline
\multicolumn{22}{c}{}\\
\multicolumn{8}{c|}{\sn-II(1p1h-1g$_{7/2}$)} & \multicolumn{7}{c|}{\sn-III(1p1h-2d$_{5/2}$)} & \multicolumn{7}{c}{\sn-IV(2p2h)}\\
\multicolumn{22}{c}{}\\
\hline
\multicolumn{22}{c}{}\\
$\mathbf{2J^+}$ & $2J_p$ & $2J_n$ & Prob.(\%) & $\pi$ WF & $\nu$ WF & Cont.(\%) & Sen. & $2J_p$ & $2J_n$ & Prob.(\%) & $\pi$ WF & $\nu$ WF & Cont.(\%) & Sen. & $2J_p$ & $2J_n$ & Prob.(\%) & $\pi$ WF & $\nu$ WF & Cont.(\%) & Sen.\\
\multicolumn{22}{c}{}\\
\hline
\hline
$\mathbf{9}$ & 9 & 0 & 88.56 & (6,4,0,1) & (10,0,0) & 63.68 & 1 & 9 & 0 & 91.01 & (6,4,0,1) & (10,0,0) & 64.90 & 1 & 9 & 0 & 91.85 & (6,4,0,1) & (10,0,0) & 57.53 & 1\\
\hline
$\mathbf{11}$ & 11 & 0 & 87.59 & (6,3,1,1) & (10,0,0) & 26.55 & 3 & 11 & 0 & 76.13 & (6,3,1,1) & (10,0,0) & 34.23 & 3 & 11 & 0 & 72.12 & (6,3,1,1) & (10,0,0) & 31.72 & 3\\
&  &  &  & (5,3,0,3) & (10,0,0) & 23.46 & 3 &  &  &  & (4,4,0,3) & (10,0,0) & 10.05 & 3 &  &  &  & (5,4,1,1) & (10,0,0) & 10.95 & 3\\
&  &  &  & (4,4,0,3) & (10,0,0) & 16.24 & 3 & 9 & 4 & 14.07 & (6,4,0,1) & (9,0,1) & 8.65 & 3 & 9 & 4 & 18.10 & (6,4,0,1) & (9,0,1) & 10.97 & 3\\
\hline
$\mathbf{13}$ & 13 & 0 & 87.87 & (4,4,0,3) & (10,0,0) & 53.21 & 3 & 13 & 0 & 85.54 & (4,4,0,3) & (10,0,0) & 49.61 & 3 & 13 & 0 & 80.99 & (4,4,0,3) & (10,0,0) & 41.96 & 3\\
&  &  &  & (5,3,0,3) & (10,0,0) & 11.61 & 3 &  &  &  & (5,3,0,3) & (10,0,0) & 12.34 & 3 &  &  &  & (5,3,0,3) & (10,0,0) & 19.17 & 3\\
\hline
$\mathbf{15}$ & 15 & 0 & 86.81 & (5,3,0,3) & (10,0,0) & 31.60 & 3 & 15 & 0 & 81.84 & (5,3,0,3) & (10,0,0) & 39.84 & 3 & 15 & 0 & 71.93 & (5,3,0,3) & (10,0,0) & 43.78 & 3\\
&  &  &  & (5,4,1,1) & (10,0,0) & 21.48 & 3 &  &  &  & (4,4,0,3) & (10,0,0) & 14.13 & 3 &  &  &  & (4,4,0,3) & (10,0,0) & 10.32 & 3\\
\hline
$\mathbf{17}$ & 17 & 0 & 87.82 & (4,4,0,3) & (10,0,0) & 61.42 & 3 & 17 & 0 & 85.73 & (4,4,0,3) & (10,0,0) & 59.71 & 3 & 17 & 0 & 55.67 & (4,4,0,3) & (10,0,0) & 39.78 & 3\\
&  &  &  & (5,3,0,3) & (10,0,0) & 12.14 & 3 &  &  &  & (5,3,0,3) & (10,0,0) & 11.73 & 3 & 9 & 10 & 14.44 & (6,4,0,1) & (9,0,1) & 9.62 &3 \\
\hline
$\mathbf{19}$ & 19 & 0 & 86.20 & (5,3,0,3) & (10,0,0) & 49.85 & 3 & 9 & 12 & 62.13 & (6,4,0,1) & (9,0,1) & 43.10 & 3 & 9 & 12 & 61.78 & (6,4,0,1) & (9,0,1) & 41.38 & 3\\
&  &  &  & (4,4,0,3) & (10,0,0) & 25.36 & 3 & 9 & 10 & 26.47 & (6,4,0,1) & (9,0,1) & 18.20 & 3 & 9 & 10 & 24.98 & (6,4,0,1) & (9,0,1) & 16.52 & 3\\
\hline
$\mathbf{21}$ & 21 & 0 & 87.36 & (4,4,0,3) & (10,0,0) & 70.13 & 3 & 9 & 14 & 51.13 & (6,4,0,1) & (9,0,1) & 36.12 & 3 & 9 & 14 & 48.69 & (6,4,0,1) & (9,0,1) & 32.82 & 3\\
&  &  &  &  &  &  &  & 9 & 12 & 39.69 & (6,4,0,1) & (9,0,1) & 27.75 & 3 & 9 & 12 & 43.07 & (6,4,0,1) & (9,0,1) & 29.03 & 3\\
\hline
$\mathbf{23}$ & 23 & 0 & 85.58 & (5,3,0,3) & (10,0,0) & 52.27 & 5 & 9 & 14 & 91.52 & (6,4,0,1) & (9,0,1) & 64.46 & 3 & 9 & 14 & 87.81 & (6,4,0,1) & (9,0,1) & 59.17 & 3\\
&  &  &  & (4,4,0,3) & (10,0,0) & 21.37 & 5 &  &  &  &  &  &  &  &  &  &  &  &  &  & \\
\hline
$\mathbf{25}$ & 25 & 0 & 86.62 & (4,4,0,3) & (10,0,0) & 67.75 & 5 & 25 & 0 & 88.30 & (4,4,0,3) & (10,0,0) & 68.66 & 3 & 25 & 0 & 83.54 & (4,4,0,3) & (10,0,0) & 66.65 & 3\\
\hline
$\mathbf{27}$ & 27 & 0 & 85.08 & (4,4,0,3) & (10,0,0) & 65.53 & 5 & 27 & 0 & 79.98 & (4,4,0,3) & (10,0,0) & 59.48 & 5 & 15 & 12 & 40.29 & (5,3,0,3) & (9,0,1) & 24.28 & 5\\
\hline
$\mathbf{29}$ & 29 & 0 & 86.44 & (4,4,0,3) & (10,0,0) & 66.49 & 5 & 29 & 0 & 88.81 & (4,4,0,3) & (10,0,0) & 68.54 & 5 & 29 & 0 & 80.15 & (4,4,0,3) & (10,0,0) & 60.6 & 5\\
&  &  &  & (4,3,1,3) & (10,0,0) & 11.32 & 7 &  &  &  & (4,3,1,3) & (10,0,0) & 11.30 & 7 &  &  &  &  &  &  & \\
\hline
$\mathbf{31}$ & 17 & 14 & 41.14 & (4,4,0,3) & (9,1,0) & 30.62 & 5 & 21 & 12 & 34.72 & (4,4,0,3) & (9,0,1) & 28.39 & 5 & 21 & 12 & 37.16 & (4,4,0,3) & (9,0,1) & 29.65 & 5\\
& 17 & 16 & 21.60 & (4,4,0,3) & (9,1,0) & 16.42 & 5 & 25 & 10 & 17.09 & (4,4,0,3) & (9,0,1) & 13.07 & 7 & 17 & 14 & 19.37 & (4,4,0,3) & (9,0,1) & 12.08 & 5\\
\hline
$\mathbf{33}$ & 21 & 14 & 33.34 & (4,4,0,3) & (9,1,0) & 27.63 & 5 & 25 & 12 & 28.28 & (4,4,0,3) & (9,0,1) & 21.84 & 7 & 21 & 12 & 28.51 & (4,4,0,3) & (9,0,1) & 23.01 & 5\\
& 17 & 16 & 16.63 & (4,4,0,3) & (9,1,0) & 12.96 & 5 & 21 & 12 & 22.07 & (4,4,0,3) & (9,0,1) & 18.14 & 5 & 25 & 12 & 20.57 & (4,4,0,3) & (9,0,1) & 16.39 & 7\\
 
& 25 & 12 & 15.24 & (4,4,0,3) & (9,1,0) & 11.76 & 5 & 21 & 14 & 17.07 & (4,4,0,3) & (9,0,1) & 13.99 & 5 & 21 & 14 & 19.46 & (4,4,0,3) & (9,0,1) & 15.49 & 5\\
 
&21  &12  & 11.52 &  (4,4,0,3) & (9,1,0)  & 9.21 & 5 & 25 & 10 & 15.03 & (4,4,0,3) & (9,0,1) & 11.4 & 7 & 19 & 14 &11.46  & (4,4,0,3) &(9,0,1)  & 6.52 & \\
\hline

$\mathbf{35}$ & 21 & 14 & 26.92 & (4,4,0,3) & (9,1,0) & 22.55 & 5 & 25 & 12 & 43.44 & (4,4,0,3) & (9,0,1) & 34.35 & 7 & 21 & 14 & 38.14 & (4,4,0,3) & (9,0,1) & 20.76 & 5\\
 
& 25 & 14 & 21.37 & (4,4,0,3) & (9,1,0) & 16.54 & 7 & 21 & 14 & 23.92 & (4,4,0,3) & (9,0,1) & 19.17 & 5 & 25 & 12 & 25.64 & (4,4,0,3) & (9,0,1) & 17.52 & 7\\
 
& 25 & 12 & 15.84 & (4,4,0,3) & (9,1,0) & 12.04 & 7 & 25 & 14 & 14.84 & (4,4,0,3) & (9,0,1) & 11.21 & 7 & 25 & 14 & 13.52 & (4,4,0,3) & (9,0,1) & 6.28 & 7\\
 
& 21 & 16 & 17.35 & (4,4,0,3) & (9,1,0) & 14.39 & 5 &  &  &  &  &  &  &  &  &  &  &  &  &  & \\
\hline
$\mathbf{37}$ & 25 & 14 & 38.78 & (4,4,0,3) & (9,1,0) & 30.52 & 7 & 25 & 12 & 38.29 & (4,4,0,3) & (9,0,1) & 29.13 & 7 & 25 & 12 & 34.16 & (4,4,0,3) & (9,0,1) & 24.78 & 7\\

& 21 & 16 & 31.76 & (4,4,0,3) & (9,1,0) & 26.48 & 5 & 25 & 14 & 23.02 & (4,4,0,3) & (9,0,1) & 18.11 & 7 & 23 & 14 & 22.69 & (4,4,0,3) & (9,0,1) & 13.54 & 7\\
\hline

$\mathbf{39}$ & 25 & 14 & 36.03 & (4,4,0,3) & (9,1,0) & 29.12 & 7 & 25 & 14 & 69.69 & (4,4,0,3) & (9,0,1) & 51.85 & 7 & 25 & 14 & 55.57 & (4,4,0,3) & (9,0,1) & 31.79 & 7\\

& 25 & 16 & 23.48 & (4,4,0,3) & (9,1,0) & 18.89 & 7 & 29 & 12 & 19.88 & (4,4,0,3) & (9,0,1) & 14.82 & 7 & 29 & 12 & 15.66 & (4,4,0,3) & (9,0,1) & 9.97 & 7\\
\hline
\hline
\end{tabular}}
\label{wavefn_snt_pos}
\end{sidewaystable}

\begin{sidewaystable}
\centering
\caption{The details of the yrast negative parity states obtained with the \sn ~interaction are presented below. The columns follow the same format and conventions as those in Table \ref{wavefn_glekpn_pos}.}
\vspace{0.5cm}
\setlength{\tabcolsep}{1pt}
\resizebox{0.8\textwidth}{!}{%
\begin{tabular}{c|ccccccc|ccccccc|ccccccc}
\hline
\multicolumn{22}{c}{}\\
\multicolumn{8}{c|}{\sn-II(1p1h-1g$_{7/2}$)} & \multicolumn{7}{c|}{\sn-III(1p1h-2d$_{5/2}$)} & \multicolumn{7}{c}{\sn-IV(2p2h)}\\
\multicolumn{22}{c}{}\\
\hline
\multicolumn{22}{c}{}\\
$\mathbf{2J^-}$ & $2J_p$ & $2J_n$ & Prob.(\%) & $\pi$ WF & $\nu$ WF & Cont.(\%) & Sen. & $2J_p$ & $2J_n$ & Prob.(\%) & $\pi$ WF & $\nu$ WF & Cont.(\%) & Sen. & $2J_p$ & $2J_n$ & Prob.(\%) & $\pi$ WF & $\nu$ WF & Cont.(\%) & Sen.\\
\multicolumn{22}{c}{}\\
\hline
\hline
$\mathbf{1}$ & 1 & 0 & 87.44 & (6,4,1,0) & (10,0,0) & 67.06 &  1 & 1 & 0 & 90.89 & (6,4,1,0) & (10,0,0) & 70.00 & 1 & 1 & 0 &  95.00& (6,4,1,0) & (10,0,0) & 67.63 & 1\\
 &  &  &  & (6,2,1,2) & (10,0,0) & 16.65 &  1 &  &  &  & (6,2,1,2) & (10,0,0) & 15.88 & 1 &  &  &  & (6,2,1,2) & (10,0,0) & 18.70 & 1\\
\hline
$\mathbf{13}$ & 13 & 0 & 87.25 & (6,3,0,2)& (10,0,0) & 47.71  & 3 & 13 & 0 & 90.10 & (6,3,0,2) & (10,0,0) & 46.05 & 3 & 13 & 0 & 89.72 & (6,3,0,2) & (10,0,0) & 46.86 & 3\\
 
&  &  &  & (5,4,0,2) &(10,0,0) & 27.76 &  3 &  &  &  & (5,4,0,2) & (10,0,0) & 31.26 & 3 &  &  &  & (5,4,0,2) & (10,0,0) & 30.94 & 3\\
\hline
$\mathbf{15}$ & 15 & 0 & 87.12 & (5,4,0,2)& (10,0,0) & 60.19  & 3 & 15 & 0 & 88.14 & (5,4,0,2) & (10,0,0) & 43.8 & 3 & 15 & 0 & 85.25 & (6,3,0,2) & (10,0,0) & 40.66 & 3\\

&  &  &  &  &  &   &  &  &  &  & (6,3,0,2) & (10,0,0) & 26.73 & 3 &  &  &  & (5,4,0,2) & (10,0,0) & 27.69 & 3\\
\hline
$\mathbf{17}$ & 17 & 0 & 87.90 & (5,4,0,2) & (10,0,0) & 71.18  & 3 & 17 & 0 & 89.32 & (5,4,0,2) & (10,0,0) & 70.89 & 3 & 17 & 0 & 87.15 & (5,4,0,2) & (10,0,0) & 68.46 & 3\\
\hline
$\mathbf{19}$ & 19 & 0 & 87.42 & (5,4,0,2) & (10,0,0)& 70.65  & 3 & 19 & 0 & 88.20 & (5,4,0,2) & (10,0,0) & 70.31 & 3 & 19 & 0 & 86.47 & (5,4,0,2) & (10,0,0) & 70.94 & 3\\
\hline
$\mathbf{21}$ & 21 & 0 & 87.43 & (5,4,0,2) & (10,0,0) & 67.44  & 3 & 21 & 0 & 89.86 & (5,4,0,2) & (10,0,0) & 69.40 & 3 & 21 & 0 & 86.98 & (5,4,0,2) & (10,0,0) & 69.16 & 3\\
\hline
$\mathbf{27}$ & 17 & 12 & 37.49 & (5,4,0,2) & (9,1,0) & 28.33  & 5 & 17 & 12 & 41.24 & (5,4,0,2) & (9,0,1) & 33.08 & 5 & 17 & 12 & 40.27 & (5,4,0,2) & (9,0,1) & 30.92 & 5\\
 
& 17 & 14 & 21.03 & (5,4,0,2) & (9,1,0) & 15.29 & 5 & 17 & 10 & 17.18 & (5,4,0,2) & (9,0,1) & 13.62 & 5 & 17 & 10 & 16.02 & (5,4,0,2) & (9,0,1) & 12.30 & 5\\

& 13 & 14 & 17.97 & (5,4,0,2) & (9,1,0)& 12.35  & 5 &  &  &  &  &  &  &  &  &  &  &  &  &  & \\
\hline
$\mathbf{29}$ & 17 & 14 & 43.68 & (5,4,0,2) & (9,1,0) & 33.91  & 5 & 17 & 12 & 40.79 & (5,4,0,2) & (9,0,1) & 33.47 & 5 & 17 & 12 & 39.10 & (5,4,0,2) & (9,0,1) & 31.61 & 5\\

& 13 & 16 & 15.83 & (5,4,0,2) & (9,1,0) & 11.62  & 5 & 19 & 12 & 20.36 & (5,4,0,2) & (9,0,1) & 16.10 & 5 & 19 & 12 & 20.01 & (5,4,0,2) & (9,0,1) & 15.38 & 5\\

& 17 & 12 & 15.43 & (5,4,0,2) & (9,1,0)& 11.52  & 5 & 19 & 10 & 12.77 & (5,4,0,2) & (9,0,1) & 9.70 & 5 & 19 & 10 & 12.23 & (5,4,0,2) & (9,0,1)  & 9.05  & \\
\hline
$\mathbf{31}$ & 17 & 14 & 38.20 & (5,4,0,2) & (9,1,0) & 30.37  & 5 & 17 & 14 & 37.65 & (5,4,0,2) & (9,0,1) & 30.93 & 5 & 17 & 14 & 37.71 & (5,4,0,2) & (9,0,1) & 29.64 & 5\\

& 19 & 14 & 20.54 & (5,4,0,2) & (9,1,0) & 15.80  & 5 & 19 & 12 & 21.35 & (5,4,0,2) & (9,0,1) & 16.72 & 5 & 19 & 12 & 20.58 & (5,4,0,2) & (9,0,1) & 15.44 & 5\\

& 19 & 12 & 14.59 & (5,4,0,2) & (9,1,0) & 10.89  & 5 & 21 & 12 & 19.69 & (5,4,0,2) & (9,0,1) & 14.86 & 5 & 21 & 12 & 17.08 & (5,4,0,2) & (9,0,1) & 12.41 & 5\\

&  &  &  &  &  &  &  & 19 & 14 & 10.41 & (5,4,0,2) & (9,0,1) & 8.21 & 5 &  19& 14  &12.12  &(5,4,0,2)  & (9,0,1) & 8.52 &5 \\
\hline
$\mathbf{33}$ & 17 & 16 & 39.96 & (5,4,0,2) & (9,1,0) & 32.08  & 5 & 21 & 12 & 44.97 & (5,4,0,2) & (9,0,1) & 34.39 & 5 & 19 & 14 & 28.33 & (5,4,0,2) & (9,1,0) & 11.16 & 5\\

& 21 & 14 & 20.77 & (5,4,0,2) & (9,1,0) & 15.65 & 5 & 19 & 14 & 36.19 & (5,4,0,2) & (9,0,1) & 29.10 & 5 &21  &12  & 17.43 & (5,4,0,2) & (9,0,1) & 9.60 & 5\\

& 19 & 14 & 18.71 & (5,4,0,2) & (9,1,0) & 14.38  & 5 & 21 & 14 & 15.76 & (5,4,0,2) & (9,0,1) & 11.72 & 5 & 17 & 16 & 23.47 & (5,4,0,2) & (9,1,0) & 18.26 & 5\\
\hline
$\mathbf{35}$ & 21 & 14 & 37.69 & (5,4,0,2) & (9,1,0) &  28.15  & 5 & 21 & 14 & 96.56 & (5,4,0,2) & (9,0,1) & 75.07 & 5 & 21 & 14 & 47.64 & (5,4,0,2) & (9,1,0) & 19.17 & 5\\

& 19 & 16 & 36.94 & (5,4,0,2) & (9,1,0) & 28.97  & 5 &  &  &  &  &  &  &  &  &  &  & (5,4,0,2) & (9,0,1) & 15.73 & 5\\

& 21 & 16 & 22.88 & (5,4,0,2)& (9,1,0) &  17.44  & 5 &  &  &  &  &  &  &  & 19 & 16 & 30.9 & (5,4,0,2) & (9,1,0) & 23.56 & 5\\

&  &  &  &  &  &  &  &  &  &  &  &  &  &  & 21 & 16 & 15.21 & (5,4,0,2) & (9,1,0) & 11.51 & 5\\
\hline
$\mathbf{37}$ & 21 & 16 & 97.57 & (5,4,0,2)& (9,1,0) & 74.80  & 5 & 23 & 14 & 47.1 & (5,3,1,2) & (9,0,1) & 33.29 & 7 & 21 & 16 & 92.37 & (5,4,0,2) & (9,1,0) & 71.4 & 5\\
&  &  &  & & &   &  & 25 & 12 & 45.83 & (5,3,1,2) & (9,0,1) & 29.93 & 7 &  &  &  &  &  &  & \\

\hline
$\mathbf{39}$ & 25 & 14 & 42.87 & (3,4,0,4) & (9,1,0)& 19.34  & 7 & 25 & 14 & 98.96 & (5,3,1,2) & (9,0,1) & 60.76 & 7 & 25 & 14 & 61.85 & (5,3,1,2) & (9,0,1) & 37.80 & 7\\

&  &  &  & (4,3,0,4) & (9,1,0)& 14.19  & 7 &  &  &  & (4,4,1,2) & (9,0,1) & 20.47 & 7 & & &  & (4,4,1,2 & (9,0,1) & 13.16 & 7\\

& 25 & 16 & 29.68 & (3,4,0,4)& (9,1,0) & 13.2  & 7 &  &  &  &  &  &  &  &  &  &  &  &  &  & \\
\hline
\hline
\end{tabular}}
\label{wavefn_snt_neg}
\end{sidewaystable}

\begin{sidewaystable}
\centering
\caption{The details of the yrast positive parity states obtained with the \gwb ~interaction are presented below. The columns follow the same format and conventions as those in Table \ref{wavefn_glekpn_pos}.}
\vspace{0.5cm}
\setlength{\tabcolsep}{1pt}
\resizebox{0.8\textwidth}{!}{%
\begin{tabular}{c|ccccccc|ccccccc|ccccccc}
\hline
\multicolumn{22}{c}{}\\
\multicolumn{8}{c|}{\gwb-II(1p1h-1g$_{7/2}$)} & \multicolumn{7}{c|}{\gwb-III(1p1h-2d$_{5/2}$)} & \multicolumn{7}{c}{\gwb-IV(2p2h)}\\
\hline
\multicolumn{22}{c}{}\\
$\mathbf{2J^+}$ & $2J_p$ & $2J_n$ & Prob.(\%) & $\pi$ WF & $\nu$ WF & Cont.(\%) & Sen. & $2J_p$ & $2J_n$ & Prob.(\%) & $\pi$ WF & $\nu$ WF & Cont.(\%) & Sen. & $2J_p$ & $2J_n$ & Prob.(\%) & $\pi$ WF & $\nu$ WF & Cont.(\%) & Sen.\\
\multicolumn{22}{c}{}\\
\hline
\hline
$\mathbf{9}$ & 9 & 0 & 96.63 & (6,4,0,1) & (10,0,0) & 77.13 & 1 & 9 & 0 & 95.81 & (6,4,0,1) & (10,0,0) & 75.16 & 1 & 9 & 0 & 94.40 & (6,4,0,1) & (10,0,0) & 73.49 & 1\\
\hline

$\mathbf{11}$ & 11 & 0 & 95.90 & (6,3,1,1) & (10,0,0) & 82.62 & 3 & 11 & 0 & 80.38 & (6,3,1,1) & (10,0,0) & 53.28 & 3 & 11 & 0 & 78.28 & (6,3,1,1) & (10,0,0) & 50.29 & 3\\
 &  &  &  &  &  &  &  &  &  &  & (5,4,1,1) & (10,0,0) & 15.30 & 3 &  &  &  & (5,4,1,1) & (10,0,0) & 16.48 & 3\\
 &  &  &  &  &  &  &  & 9 & 4 & 14.53 & (6,4,0,1) & (9,0,1) & 10.53 & 3 & 9 & 4 & 14.62 & (6,4,0,1) & (9,0,1) & 10.20 & 3\\
 
\hline
$\mathbf{13}$ & 13 & 0 & 95.86 & (5,4,1,1) & (10,0,0) & 46.03 & 3 & 13 & 0 & 84.81 & (6,3,1,1) & (10,0,0) & 59.71 & 3 & 13 & 0 & 83.12 & (6,3,1,1) & (10,0,0) & 60.04 & 3\\
 &  &  &  & (6,3,1,1) & (10,0,0) & 29.47 & 3 &  &  &  &  &  &  &  &  &  &  &  &  &  & \\
\hline
$\mathbf{15}$ & 15 & 0 & 95.65 & (5,4,1,1) & (10,0,0) & 70.58 & 3 & 15 & 0 & 92.80 & (5,4,1,1) & (10,0,0) & 65.12 & 3 & 9 & 8 & 32.71 & (6,4,0,1) & (9,0,1) & 23.77 & 3\\
 &  &  &  &  &  &  &  &  &  &  &  &  &  &  & 9 & 10 & 26.79 & (6,4,0,1) & (9,0,1) & 19.69 & 3\\
 &  &  &  &  &  &  &  &  &  &  &  &  &  &  & 15 & 0 & 17.54 & (5,4,1,1) & (10,0,0) & 10.66 & 3\\
\hline
$\mathbf{17}$ & 17 & 0 & 93.61 & (4,4,0,3) & (10,0,0) & 67.50 & 3 & 9 & 10 & 45.00 & (6,4,0,1) & (9,0,1) & 34.12 & 3 & 9 & 10 & 42.40 & (6,4,0,1) & (9,0,1) & 31.31 & 3\\

&  &  &  & (5,3,0,3) & (10,0,0) & 10.83 & 3 & 9 & 12 & 27.25 & (6,4,0,1) & (9,0,1) & 20.53 & 3 & 9 & 12 & 30.81 & (6,4,0,1) & (9,0,1) & 22.39 & 3\\

&  &  &  &  &  &  &  & 9 & 8 & 17.8 & (6,4,0,1) & (9,0,1) & 13.56 & 3 & 9 & 8 & 15.68 & (6,4,0,1) & (9,0,1) & 11.53 & 3\\
\hline
$\mathbf{19}$ & 19 & 0 & 93.60 & (5,3,0,3) & (10,0,0) & 42.37 & 5 & 9 & 12 & 62.68 & (6,4,0,1) & (9,0,1) & 48.33 & 3 & 9 & 12 & 61.28 & (6,4,0,1) & (9,0,1) & 46.02 & 3\\
 &  &  &  & (4,4,0,3) & (10,0,0) & 38.94 & 3 & 9 & 10 & 29.62 & (6,4,0,1) & (9,0,1) & 22.81 & 3 & 9 & 10 & 28.67 & (6,4,0,1) & (9,0,1) & 21.44 & 3\\
\hline
$\mathbf{21}$ & 21 & 0 & 93.13 & (4,4,0,3) & (10,0,0) & 74.81 & 3 & 9 & 12 & 49.92 & (6,4,0,1) & (9,0,1) & 39.19 & 3 & 9 & 12 & 46.68 & (6,4,0,1) & (9,0,1) & 35.72 & 3\\
 &  &  &  &  &  &  &  & 9 & 14 & 45.10 & (6,4,0,1) & (9,0,1) & 35.18 & 3 & 9 & 14 & 46.51 & (6,4,0,1) & (9,0,1) & 35.26 & 3\\
\hline
$\mathbf{23}$ & 23 & 0 & 93.30 & (5,3,0,3) & (10,0,0) & 47.56 & 5 & 9 & 14 & 94.70 & (6,4,0,1) & (9,0,1) & 75.70 & 3 & 9 & 14 & 92.34 & (6,4,0,1) & (9,0,1) & 72.20 & 3\\
 &  &  &  & (4,4,0,3) & (10,0,0) & 32.43 & 5 &  &  &  &  &  &  &  &  &  &  &  &  &  & \\
\hline
$\mathbf{25}$ & 25 & 0 & 92.51 & (4,4,0,3) & (10,0,0) & 72.14 & 5 & 25 & 0 & 91.68 & (4,4,0,3) & (10,0,0) & 70.57 & 5 & 25 & 0 & 87.08 & (4,4,0,3) & (10,0,0) & 70.61 & 5\\

&  &  &  & (4,3,1,3) & (10,0,0) & 10.24 & 5 &  &  &  & (4,3,1,3) & (10,0,0) & 10.19 & 5 &  &  &  &  &  &  & \\
\hline

$\mathbf{27}$ & 27 & 0 & 90.88 & (4,4,0,3) & (10,0,0) & 70.31 & 5 & 27 & 0 & 90.65 & (4,4,0,3) & (10,0,0) & 68.37 & 5 & 27 & 0 & 82.97 & (4,4,0,3) & (10,0,0) & 64.99 & 5\\
 
&  &  &  & (4,3,1,3) & (10,0,0) & 11.63 & 5 &  &  &  & (4,3,1,3) & (10,0,0) & 11.46 & 5 &  &  &  &  &  &  & \\
\hline

$\mathbf{29}$ & 29 & 0 & 92.28 & (4,4,0,3) & (10,0,0) & 70.57 & 5 & 29 & 0 & 92.24 & (4,4,0,3) & (10,0,0) & 71.05 & 5 & 29 & 0 & 86.33 & (4,4,0,3) & (10,0,0) & 67.24 & 5\\

&  &  &  & (4,3,1,3) & (10,0,0) & 13.12 & 7 &  &  &  & (4,3,1,3) & (10,0,0) & 13.01 & 7 &  &  &  & (4,3,1,3)  & (10,0,0) & 11.54 & 5\\
\hline

$\mathbf{31}$ & 17 & 14 & 42.48 & (4,4,0,3) & (9,1,0) & 32.37 & 5 & 21 & 12 & 40.2 & (4,4,0,3) & (9,0,1) & 33.46 & 5 & 21 & 12 & 45.16 & (4,4,0,3) & (9,0,1) & 37.54 & 5\\

& 17 & 16 & 17.22 & (4,4,0,3) & (9,1,0) & 13.02 & 5 & 25 & 10 & 17.5 & (4,4,0,3) & (9,0,1) & 13.56 & 7 & 17 & 14 & 27.9 & (4,4,0,3) & (9,0,1) & 20.69 & 5\\

& 21 & 14 & 12.01 & (4,4,0,3) & (9,1,0) & 9.73 & 5 &  &  &  &  &  &  &  &  &  &  &  &  &  & \\

& 21 & 12 & 10.41 & (4,4,0,3) & (9,1,0) & 8.61 & 5 &  &  &  &  &  &  &  &  &  &  &  &  &  & \\

\hline
$\mathbf{33}$ & 21 & 14 & 35.88 & (4,4,0,3) & (9,1,0) & 29.75 & 5 & 25 & 12 & 27.48 & (4,4,0,3) & (9,0,1) & 21.35 & 7 & 21 & 12 & 32.02 & (4,4,0,3) & (9,0,1) & 26.26 & 5\\

& 25 & 12 & 16.48 & (4,4,0,3) & (9,1,0) & 12.63 & 7 & 21 & 12 & 24.36 & (4,4,0,3) & (9,0,1) & 20.26 & 5 & 25 & 12 & 19.72 & (4,4,0,3) & (9,0,1) & 16.14 & 7\\

& 17 & 16 & 14.20 & (4,4,0,3) & (9,1,0) & 11.15 & 5 & 21 & 14 & 18.67 & (4,4,0,3) & (9,0,1) & 15.46 & 5 & 21 & 14 & 18.72 & (4,4,0,3) & (9,0,1) & 15.20 & 5\\

& 21 & 12 & 12.46 & (4,4,0,3) & (9,1,0) & 10.00 & 5 & 25 & 10 & 14.79 & (4,4,0,3) & (9,0,1) & 11.35 & 7 &  &  &  &  &  &  & \\

\hline
$\mathbf{35}$ & 21 & 14 & 26.63 & (4,4,0,3) & (9,1,0) & 21.99 & 7 & 25 & 12 & 44.17 & (4,4,0,3) & (9,0,1) & 35.13 & 7 & 21 & 14 & 44.88 & (4,4,0,3) & (9,0,1) & 32.53 & 5\\

& 25 & 14 & 23.66 & (4,4,0,3) & (9,1,0) & 18.07 & 7 & 21 & 14 & 26.39 & (4,4,0,3) & (9,0,1) & 21.28 & 5 & 25 & 12 & 31.12 & (4,4,0,3) & (9,0,1) & 24.28 & 7\\

& 25 & 12 & 18.39 & (4,4,0,3) & (9,1,0) & 13.79 & 7 & 25 & 14 & 14.78 & (4,4,0,3) & (9,0,1) & 11.25 & 7 &  &  &  &  &  &  & \\

& 21 & 16 & 16.59 & (4,4,0,3) & (9,1,0) & 13.68 & 5 &  &  &  &  &  &  &  &  &  &  &  &  &  & \\
\hline
$\mathbf{37}$ & 25 & 14 & 44.73 & (4,4,0,3) & (9,1,0) & 34.64 & 7 & 25 & 12 & 46.37 & (4,4,0,3) & (9,0,1) & 37.52 & 7 & 25 & 12 & 43.88 & (4,4,0,3) & (9,0,1) & 34.67 & 7\\

& 21 & 16 & 25.67 & (4,4,0,3) & (9,1,0) & 20.70 & 5 & 25 & 14 & 33.02 & (4,4,0,3) & (9,0,1) & 26.69 & 7 & 25 & 14 & 22.42 & (4,4,0,3) & (9,0,1) & 18.28 & 7\\
 
& 25 & 16 & 11.36 & (4,4,0,3) & (9,1,0) & 8.32 & 7 &  &  &  &  &  &  &  & 23 & 14 & 13.26 & (4,4,0,3) & (9,0,1) & 6.28 & 7\\
 
& 25 & 12 & 10.71 & (4,4,0,3) & (9,1,0) & 7.89 & 7 &  &  &  &  &  &  &  &  &  &  &  &  &  & \\
\hline
$\mathbf{39}$ & 25 & 14 & 46.24 & (4,4,0,3) & (9,1,0) & 36.86 & 7 & 25 & 14 & 84.62 & (4,4,0,3) & (9,0,1) & 66.84 & 7  & 25 & 14 & 79.41 & (4,4,0,3) & (9,0,1) & 60.07 & 7\\
 
& 25 & 16 & 29.45 & (4,4,0,3) & (9,1,0) & 23.25 & 7 &  &  &  &  &  &  &  & 29 & 12 & 11.49 & (4,4,0,3) & (9,0,1) & 8.52 & 7\\
\hline
\hline
\end{tabular}}
\label{wavefn_gwbxg_pos}
\end{sidewaystable}

\begin{sidewaystable}
\centering
\caption{The details of the yrast negative parity states obtained with the \gwb ~interaction are presented below. The columns follow the same format and conventions as those in Table \ref{wavefn_glekpn_pos}. }
\vspace{0.5cm}
\setlength{\tabcolsep}{1pt}
\resizebox{0.8\textwidth}{!}{%
\begin{tabular}{c|ccccccc|ccccccc|ccccccc}
\hline
\multicolumn{22}{c}{}\\
\multicolumn{8}{c|}{\gwb-II(1p1h-1g$_{7/2}$)} & \multicolumn{7}{c|}{\gwb-III(1p1h-2d$_{5/2}$)} & \multicolumn{7}{c}{\gwb-IV(2p2h)}\\
\multicolumn{22}{c}{}\\
\hline
\multicolumn{22}{c}{}\\
$\mathbf{2J^-}$ & $2J_p$ & $2J_n$ & Prob.(\%) & $\pi$ WF & $\nu$ WF & Cont.(\%) & Sen. & $2J_p$ & $2J_n$ & Prob.(\%) & $\pi$ WF & $\nu$ WF & Cont.(\%) & Sen. & $2J_p$ & $2J_n$ & Prob.(\%) & $\pi$ WF & $\nu$ WF & Cont.(\%) & Sen.\\
\multicolumn{22}{c}{}\\
\hline
\hline
$\mathbf{1}$ & 1 & 0 & 97.75 & (6,4,1,0) & (10,0,0) & 82.75 & 1 & 1 & 0 & 98.16 & (6,4,1,0) & (10,0,0) & 82.7 & 1 & 1 & 0 & 98.17 & (6,4,1,0) & (10,0,0) & 81.11 & 1\\
 &  &  &  & (6,2,1,2) & (10,0,0) & 10.99 & 1 &  &  &  & (6,2,1,2) & (10,0,0) & 11.27 & 1 &  &  &  & (6,2,1,2) & (10,0,0) &  12.37& 1 \\
\hline

$\mathbf{13}$ & 13 & 0 & 95.45 & (5,4,0,2) & (10,0,0) & 42.44 & 3 & 13 & 0 & 94.89 & (5,4,0,2) & (10,0,0) & 41.97 & 3 & 13 & 0 & 92.18 & (6,3,0,2) & (10,0,0) & 41.97 & 3\\

&  &  &  & (6,3,0,2) & (10,0,0) & 41.62 & 3 &  &  &  & (6,3,0,2) & (10,0,0) & 40.84 & 3 &  &  &  & (5,4,0,2) & (10,0,0) & 38.91 & 3\\
\hline

$\mathbf{15}$ & 15 & 0 & 94.27 & (5,4,0,2) & (10,0,0) & 60.02 & 3 & 15 & 0 & 92.77 & (5,4,0,2) & (10,0,0) & 56.49 & 3 & 15 & 0 & 88.44 & (5,4,0,2) & (10,0,0) & 40.66 & 3\\

&  &  &  & (6,3,0,2) & (10,0,0) & 15.60 & 3 &  &  &  & (6,3,0,2) & (10,0,0) & 16.57 & 3 &  &  &  & (6,3,0,2) & (10,0,0) & 27.32 & 3\\
\hline

$\mathbf{17}$ & 17 & 0 & 94.92 & (5,4,0,2) & (10,0,0) & 78.74 & 3 & 17 & 0 & 93.38 & (5,4,0,2) & (10,0,0) & 76.03 & 3 & 17 & 0 & 90.07 & (5,4,0,2) & (10,0,0) & 72.40 & 3\\

\hline
$\mathbf{19}$ & 19 & 0 & 94.86 & (5,4,0,2) & (10,0,0) & 78.59 & 3 & 19 & 0 & 92.82 & (5,4,0,2) & (10,0,0) & 75.53 & 3 & 19 & 0 & 89.75 & (5,4,0,2) & (10,0,0) & 74.23 & 3\\
\hline

$\mathbf{21}$ & 21 & 0 & 94.36 & (5,4,0,2) & (10,0,0) & 75.71 & 3 & 21 & 0 & 93.57 & (5,4,0,2) & (10,0,0) & 74.01 & 3 & 21 & 0 & 89.75 & (5,4,0,2) & (10,0,0) & 71.25 & 3\\
\hline

$\mathbf{27}$ & 17 & 12 & 40.42 & (5,4,0,2) & (9,1,0) & 31.17 & 5 & 17 & 12 & 40.49 & (5,4,0,2) & (9,0,1) & 33.63 & 5 & 17 & 12 & 40.44 & (5,4,0,2) & (9,0,1) & 32.31 & 5\\

& 17 & 14 & 19.33 & (5,4,0,2) & (9,1,0) & 14.07 & 5 & 17 & 10 & 22.11 & (5,4,0,2) & (9,0,1) & 18.50 & 5 & 17 & 10 & 19.42 & (5,4,0,2) & (9,0,1) & 15.70 & 5 \\

& 13 & 14 & 17.18 & (5,4,0,2) & (9,1,0) & 11.40 & 5 &  &  &  &  &  &  &  &  &  &  &  &  &  & \\
\hline

$\mathbf{29}$ & 17 & 14 & 45.26 & (5,4,0,2) & (9,1,0) & 35.47 & 5 & 17 & 12 & 43.99 & (5,4,0,2) & (9,0,1) & 37.73 & 5 & 17 & 12 & 41.06 & (5,4,0,2) & (9,0,1) & 34.6 & 5\\

& 17 & 12 & 19.77 & (5,4,0,2) & (9,1,0) & 15.21 & 5 & 19 & 12 & 18.38 & (5,4,0,2) & (9,0,1) & 15.05 & 5 & 19 & 12 & 18.33 & (5,4,0,2) & (9,0,1) & 14.54 & 5\\

& 13 & 16 & 10.63 & (5,4,0,2) & (9,1,0) & 7.02 & 5 & 19 & 10 & 14.44 & (5,4,0,2) & (9,0,1) & 11.59 & 5 & 19 & 10 & 13.19 & (5,4,0,2) & (9,0,1) & 10.26 & 5\\
\hline

$\mathbf{31}$ & 17 & 14 & 43.05 & (5,4,0,2) & (9,1,0) & 35.49 & 5 & 17 & 14 & 39.56 & (5,4,0,2) & (9,0,1) & 34.09 & 5 & 17 & 14 & 40.1 & (5,4,0,2) & (9,0,1) & 33.69 & 5\\

& 19 & 14 & 18.64 & (5,4,0,2) & (9,1,0) & 14.68 & 5 & 19 & 12 & 26.54 & (5,4,0,2) & (9,0,1) & 21.82 & 5 & 19 & 12 & 23.14 & (5,4,0,2) & (9,0,1) & 18.45 & 5\\

& 19 & 12 & 14.80 & (5,4,0,2) & (9,1,0) & 11.52 & 5 & 21 & 12 & 15.65 & (5,4,0,2) & (9,0,1) & 12.14 & 5 & 21 & 12 & 15.23 & (5,4,0,2) & (9,0,1) & 11.42 & 5\\
 
&  &  &  &  &  &  &  &  &  &  &  &  &  &  & 19 & 14 & 10.04 & (5,4,0,2) & (9,0,1) & 7.52 & 5\\
 \hline
$\mathbf{33}$ & 17 & 16 & 44.35 & (5,4,0,2) & (9,1,0) & 36.88 & 5 & 21 & 12 & 49.40 & (5,4,0,2) & (9,0,1) & 39.46 & 5 & 19 & 14 & 42.92 & (5,4,0,2) & (9,0,1) & 33.90 & 5\\

& 19 & 14 & 23.5 & (5,4,0,2) & (9,1,0) & 18.78 & 5 & 19 & 14 & 41.84 & (5,4,0,2) & (9,0,1) & 35.08 & 5 & 21 & 12 & 39.51 & (5,4,0,2) & (9,0,1) & 30.25 & 5\\

& 21 & 14 & 17.38 & (5,4,0,2) & (9,1,0) & 13.37 & 5 &  &  &  &  &  &  &  &  &  &  &  &  &  & \\
\hline
$\mathbf{35}$ & 19 & 16 & 47.70 & (5,4,0,2) & (9,1,0) & 38.76 & 5 & 21 & 14 & 97.27 & (5,4,0,2) & (9,0,1) & 79.16 & 5 & 21 & 14 & 91.66 & (5,4,0,2) & (9,0,1) & 72.36 & 5\\

& 21 & 14 & 42.29 & (5,4,0,2) & (9,1,0) & 32.42 & 5 &  &  &  &  &  &  &  &  &  &  &  &  &  & \\
\hline

$\mathbf{37}$ & 21 & 16 & 97.73 & (5,4,0,2) & (9,1,0) & 77.42 & 5 & 25 & 12 & 55.39 & (5,3,1,2) & (9,0,1) & 34.06 & 7 & 21 & 16 & 92.61 & (5,4,0,2) & (9,1,0) & 71.95 & 5\\
 
&  &  &  &  &  &  &  &  &  &  & (4,4,1,2) & (9,0,1) & 14.81 & 7 &  &  &  &  &  &  & \\
 
&  &  &  &  &  &  &  & 23 & 14 & 44.13 & (5,3,1,2) & (9,0,1) & 29.02 & 7 &  &  &  &  &  &  & \\
 
&  &  &  &  &  &  &  &  &  &  & (4,4,1,2) & (9,0,1) & 10.59 & 7 &  &  & &&&&  \\
\hline

$\mathbf{39}$ & 23 & 16 & 60.07 & (5,3,1,2) & (9,1,0) & 31.43 & 7 & 25 & 14 & 99.59 & (5,3,1,2) & (9,0,1) & 55.96 & 7 & 25 & 14 & 92.74 & (5,3,1,2) & (9,0,1) & 47.07 & 7\\

&  &  &  & (4,4,1,2) & (9,1,0) & 21.42 & 7 &  &  &  & (4,4,1,2) & (9,0,1) & 32.79 & 7 &  &  &  & (4,4,1,2) & (9,0,1) & 32.65 & 7\\

& 25 & 14 & 38.85 & (5,3,1,2) & (9,1,0) & 21.83 & 7 &  &  &  &  &  &  &  &  &  &  &  &  &  & \\

&  &  &  & (4,4,1,2) & (9,1,0) & 11.57 & 7 &  &  &  &  &  &  &  &  &  &  &  &  &  & \\
\hline
\hline
\end{tabular}}
\label{wavefn_gwbxg_neg}
\end{sidewaystable}


\subsection{\label{occupation} Occupation plot }

\begin{figure}[htbp]
    \centering

    \begin{subfigure}[b]{0.48\textwidth}
        \centering
        \includegraphics[width=\linewidth]{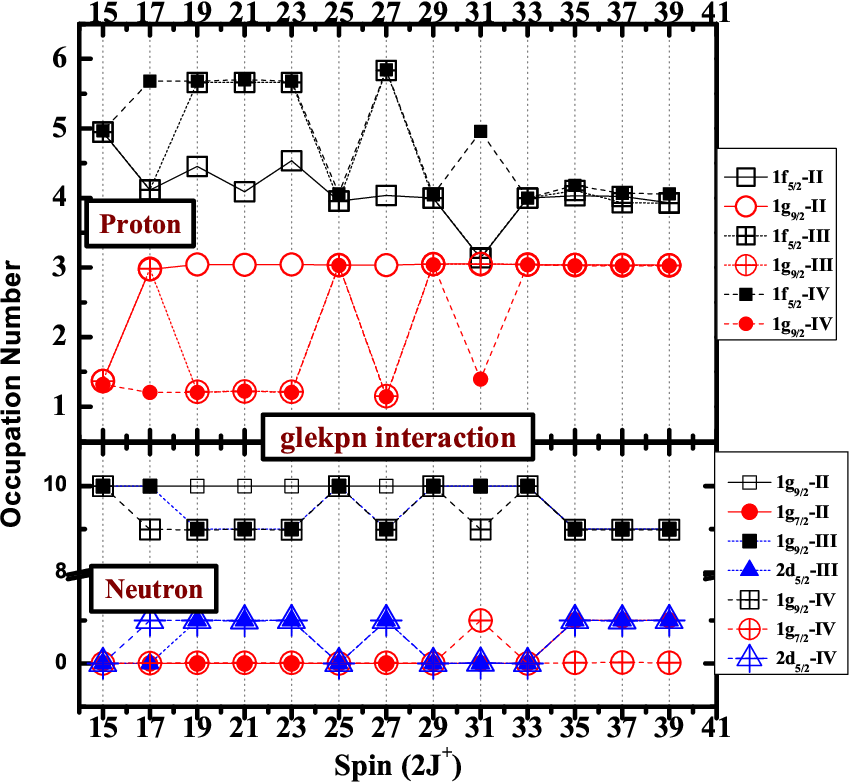}
        \caption{For different truncations with \gle ~interactions}
        \label{fig:pos_gle}
    \end{subfigure}
    \hfill
    \begin{subfigure}[b]{0.48\textwidth}
        \centering
        \includegraphics[width=\linewidth]{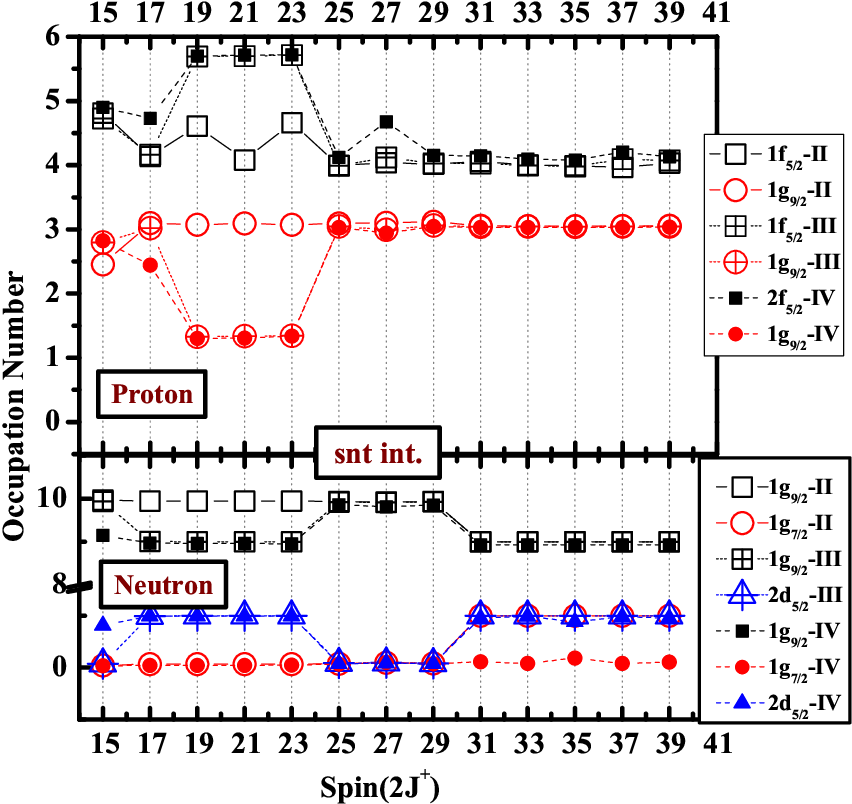}
        \caption{For different truncations with \sn ~interactions}
        \label{fig:pos_snt}
    \end{subfigure}

    \vspace{0.25cm}

    \begin{subfigure}[b]{0.48\textwidth}
        \centering
        \includegraphics[width=\linewidth]{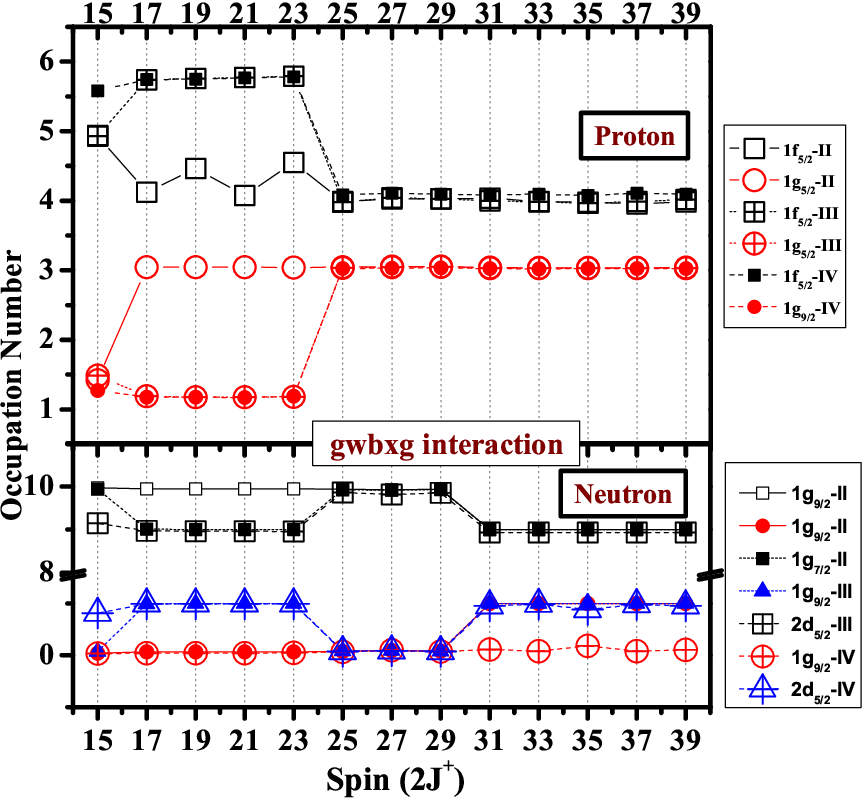}
        \caption{For different truncations with \gwb ~interactions}
        \label{fig:pos_gwbxg}
    \end{subfigure}

    \vspace{-0.15cm}

    \caption{Occupation probabilities for yrast positive parity states obtained using \gle, \sn, and \gwb ~interactions for different truncations.}
    \label{occupation_pos_parity}
\end{figure}

\begin{figure}[htbp]
    \centering

    \begin{subfigure}[b]{0.48\textwidth}
        \centering
        \includegraphics[width=\linewidth]{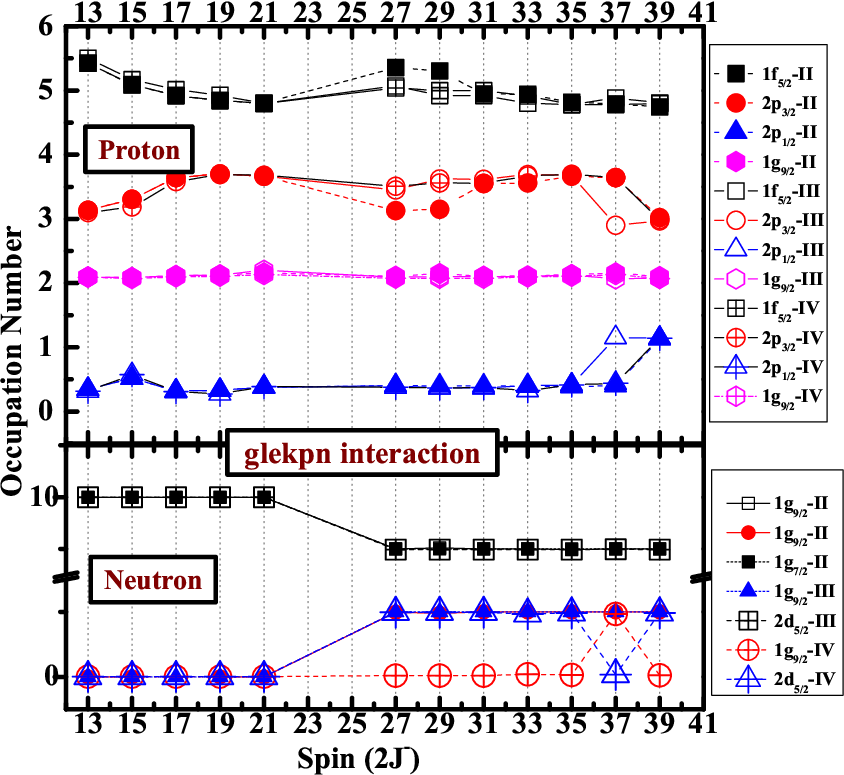}
        \caption{For different truncations with \gle ~interaction}
        \label{fig:neg_gle}
    \end{subfigure}
    \hfill
    \begin{subfigure}[b]{0.48\textwidth}
        \centering
        \includegraphics[width=\linewidth]{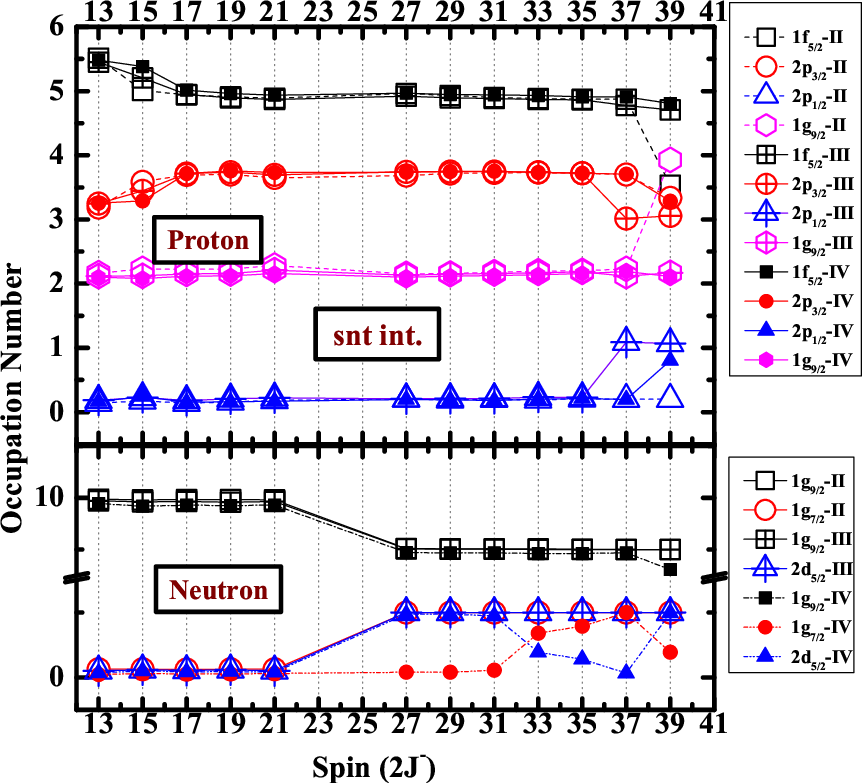}
        \caption{For different truncations with \sn ~interaction}
        \label{fig:neg_snt}
    \end{subfigure}

    \vspace{0.25cm}

    \begin{subfigure}[b]{0.48\textwidth}
        \centering
        \includegraphics[width=\linewidth]{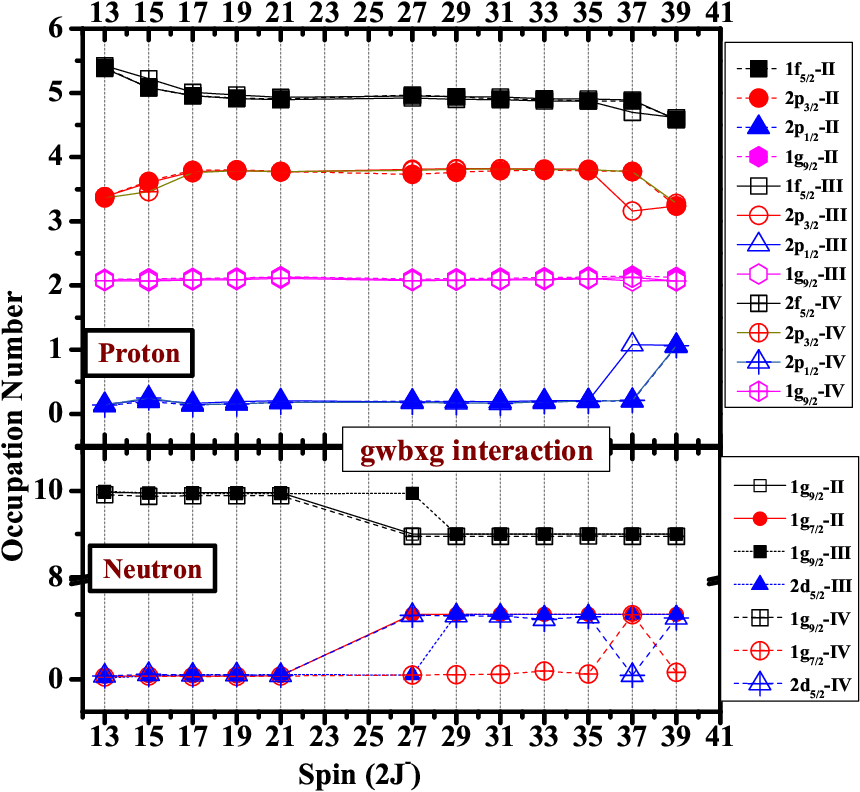}
        \caption{For different truncations with \gwb ~interaction}
        \label{fig:neg_gwbxg}
    \end{subfigure}

    \vspace{-0.15cm}

    \caption{Occupation probabilities for yrast negative parity states obtained using \gle, \sn, and \gwb ~interactions for different truncations.}
    \label{occupation_neg_parity}
\end{figure}

 The evolution of average proton and neutron occupancies with spin has been examined for the cross shell interactions, considering n-excitation (Fig. \ref{occupation_pos_parity} and Fig. \ref{occupation_neg_parity}). For positive parity states, the occupancies of the $\pi 2p_{3/2}$ and $\pi 2p_{1/2}$ orbitals remain nearly constant across the spin range and for all truncation schemes; hence, they are not shown in the figure. In contrast, the $\pi 1f_{5/2}$ and $\pi 1g_{9/2}$ orbitals exhibit noticeable variations depending on both spin and truncation. A significant difference arises for the $1p$–$1h$ (restriction no. II - $\nu 1g_{7/2}$) truncation, which can be attributed to the relatively high single particle energy of the $\nu 1g_{7/2}$ orbital as pointed out earlier. It suppresses neutron excitation across the $N=50$ shell gap. In the spin range $17/2^{+}$–$23/2^{+}$, the states are predominantly proton driven in this truncation, as reflected by a decrease in the $\pi 1f_{5/2}$ occupancy indicating proton pair breaking  and a corresponding increase in the $\pi 1g_{9/2}$ occupancy signifying the promotion of these two protons. In contrast, for the $1p$–$1h$ (restriction no. III - $2d_{5/2}$) and $2p$–$2h$ truncations, a non-zero occupancy of the $\nu 2d_{5/2}$ orbital clearly indicates the onset of neutron core excitation through breaking of the $\nu 1g_{9/2}$ pair. A noticeable change in both neutron and proton occupancies is observed in the $25/2^{+}$–$29/2^{+}$ region, suggesting that proton pair breaking and neutron core excitation become comparable in this energy range, with proton driven configurations remaining slightly more favorable. The subsequent states are generally generated through a combined mechanism involving both proton excitations and neutron core excitations, as clearly reflected in the evolution of the occupation numbers.

For the yrast negative parity states, a similar trend in the average occupation numbers is observed across all truncation schemes. The $\pi 1f_{5/2}$ occupancy shows a slight decrease around the $15/2^{-}$ state, accompanied by a complementary increase in the $\pi 2p_{3/2}$ orbital, after which both remain nearly constant with increasing spin. The occupancies of the $\pi 2p_{1/2}$ and $\pi 1g_{9/2}$ orbitals also exhibit only minor variations, indicating their relatively stable contribution in this spin range. Neutron core excitation becomes evident from the occupancy of the $\nu 1g_{7/2}$ or $\nu 2d_{5/2}$ orbitals, where non-zero contributions appear from around the $27/2^{-}$ state. A noticeable change in the proton occupancies is observed near the $37/2^{-}$ state, suggesting their increasing role in generating higher angular momentum. This behavior, together with the onset of neutron excitation, indicates that the high spin states are formed through a combined mechanism involving proton alignment and breaking of the $\nu 1g_{9/2}$ neutron pair.
\subsection{\label{Transition} Transition Probability }
The SM obtained transition probabilities $B(E2)$ and $B(M1)$ are compared with the corresponding available experimentally measured  results for the yrast states (Tables~\ref{be2} and~\ref{bm1}). For the cross shell interactions, calculations are performed using two $1p1h$ truncations, corresponding to excitations into the $1g_{7/2}$ and $2d_{5/2}$ orbitals. In most cases, the proton and neutron occupancies obtained with the $2p2h$ restriction are very similar to those obtained with the $1p1h(2d_{5/2})$ restriction. Hence, comparable transition strengths are expected. Therefore, the $2p2h$ transition strengths are not included in the tables. The $0p0h$ configuration is considered only for the $\ju$ interaction. For the $E2$ transitions, standard effective charges, $e_p^{\rm eff}=1.5e$ and $e_n^{\rm eff}=0.5e$, are adopted. For the $M1$ transitions, bare-nucleon $g$ factors, $g_p^s=5.586$, $g_n^s=-3.826$, $g_p^l=1.0$, and $g_n^l=0$, are used.

The $B(E2;3/2^{-}\rightarrow1/2^{-})$ transition is well reproduced by both restrictions of the $\gle$ interaction, whereas $\ju$ slightly underestimates the experimental value and the $\sn$ and $\gwb$ interactions show larger discrepancies. For the $5/2^{-}\rightarrow1/2^{-}$ transition, relatively good agreement is obtained only with the $\ju$ interaction, while the other interactions underestimate the experimental strength. None of the interactions reproduces the experimental $B(E2;15/2^-\rightarrow13/2^-)$ value in good agreement; all calculations substantially overestimate the transition strength. For the $B(E2;17/2^-\rightarrow13/2^-)$ transition, the agreement is mixed: $\ju$ and both $\gle$ restrictions underestimate the experimental strength, whereas $\sn$-II, $\gwb$-I, and $\gwb$-II overestimate it. In contrast, the $B(E2;19/2^-\rightarrow17/2^-)$ strength is overestimated by all interactions. Overall, the $\ju$ interaction with the $0p0h$ configuration gives a relatively good description of the negative parity transition strengths, while the $1p1h$ configurations of $\gle$ provide improved agreement for selected transitions. This comparatively better agreement with $\ju$ is physically consistent with the predominantly proton excitation character of these negative parity states.

For the low lying positive parity states up to $13/2^{+}$, the $\ju$ interaction again provides a reasonable description of the transition strengths. A notable exception is the $13/2^{+}\rightarrow9/2^{+}$ transition, for which the $1p1h(2d_{5/2})$ truncation gives a markedly closer reproduction of the experimental strength than the $1p1h(1g_{7/2})$ truncation, despite both the initial and final states being dominated by proton excitation (Table~\ref{transition_be2}). A similar preference for the $2d_{5/2}$ restriction is observed for the $17/2^{+}\rightarrow13/2^{+}$ transition across the cross shell interactions. For the $23/2^{+}\rightarrow21/2^{+}$ transition, the $2d_{5/2}$ restriction gives closer agreement with experiment than the $1g_{7/2}$ restriction for $\gle$, although the calculated strength remains overestimated.  The yrast $23/2^+$ and $21/2^+$ states are primarily associated with neutron core excitation in the $2d_{5/2}$ space, whereas proton excitation dominates in the $1g_{7/2}$ space.  This difference in the dominant wavefunction components naturally leads to different transition matrix elements and, consequently, different calculated transition strengths. The E2 strength for $25/2^{+} \to 21/2^{+}$  is best reproduced by \ju, while among the cross shell interactions the 1g$_{7/2}$ restriction consistently outperforms 2d$_{5/2}$ . This is likely because both states are generated primarily from proton excitation; the greater neutron configuration mixing introduced by the 2d$_{5/2}$ restriction is not representative of the dominant structure and degrades the prediction. For the $29/2^{+}\rightarrow27/2^{+}$ transition, only $\ju$ reproduces the experimental strength reasonably well, while all cross shell calculations underestimate the transition strength, with the magnitude of the discrepancy depending strongly on the adopted restriction. None of the interactions and restrictions reproduce the transition rates for the higher positive parity states beyond this point. Overall, the relative performance of the two cross shell restrictions is strongly transition dependent. For several transitions, the calculated $B(E2)$ values obtained with the $1g_{7/2}$ and $2d_{5/2}$ restrictions are very similar, indicating a limited sensitivity to the choice of the excited orbital. In contrast, substantial differences arise for selected transitions, particularly when the corresponding wavefunctions acquire different proton and neutron core excitation components. The $2d_{5/2}$ restriction gives closer agreement for the $13/2^+\rightarrow9/2^+$, $17/2^+\rightarrow13/2^+$, and $23/2^+\rightarrow21/2^+$ transitions, whereas the $1g_{7/2}$ restriction gives closer agreement for the $25/2^+\rightarrow21/2^+$ and $29/2^+\rightarrow27/2^+$ transitions. Thus, the relative performance of the two restrictions appears to be governed by the detailed configuration mixing and proton--neutron structure of the initial and final states.

In case of M1 transition strength (Table \ref{transition_bm1}), no interaction provides consistent agreement with experiment across the negative parity states. For the sequence of 15/2$^{-}$, 17/2$^{-}$ and 19/2$^{-}$ and 21/2$^{-}$, the \ju ~gives the closest prediction for $15/2^{-} \to 13/2^{-}$). However, other predicted values are mostly underestimated for \ju ~interaction and overestimated for other cross shell interactions. For the positive parity states, the \ju ~interaction generally gives poor agreement, with discrepancies of one to three orders of magnitude for several transitions. For the $19/2^{+} \to 17/2^{+}$ transition, the best result is achieved with the \gwb ~with restriction 1p1h (restriction no. III - $\nu 2d_{5/2}$), where both initial and final states are dominated by neutron core excitation. The other interactions and restrictions instead produce wavefunctions dominated by proton excitation, resulting in transition strengths that are approximately three orders of magnitude smaller than the experimental value. A similar advantage for the 2d$_{5/2}$, restriction is seen at $21/2^{+} \to 19/2^{+}$, where both \gle-III and \gwb-III outperform their respective 1g$_{7/2}$ counterparts. At $23/2^{+} \to 21/2^{+}$, however, the pattern is mixed rather than uniform: the 2d$_{5/2}$, restriction is closer to experiment for \gle, while the 1g$_{7/2}$ restriction is closer for \gwb. For the higher sequence from $25/2^{+} \to 23/2^{+}$ and $29/2^{+} \to 27/2^{+}$,  the 1p1h-1g$_{7/2}$ restriction generally gives relatively better agreement. This may be associated with the smaller neutron core mixing obtained with the $1g_{7/2}$ restriction, owing to the relative position of the $1g_{7/2}$ orbital. This is clearest at $25/2^{+} \to 23/2^{+}$  and $29/2^{+} \to 27/2^{+}$ where 1g$_{7/2}$ outperforms 2d$_{5/2}$, for both glekpn and gwbxg. This advantage is weaker and less consistent at $27/2^{+} \to 25/2^{+}$. However, at $31/2^{+} \to 29/2^{+}$, both restrictions fail comparably badly. Overall, the calculated strengths remain poorly reproduced for most of these higher-spin transitions, irrespective of the adopted restriction. The strong sensitivity of $B(M1)$ values to the detailed spin distribution among the valence orbitals may account for the difficulty in reproducing the experimental strengths, even when neutron core excitations are included. Overall, no uniform preference for either cross shell restriction is observed in the $B(M1)$ transition strengths. The $2d_{5/2}$ restriction gives closer agreement for the $21/2^+\rightarrow19/2^+$ and $23/2^+\rightarrow21/2^+$ transitions, whereas the $1g_{7/2}$ restriction provides relatively better agreement for the $25/2^+\rightarrow23/2^+$ and $29/2^+\rightarrow27/2^+$ transitions. This transition dependent behavior reflects the sensitivity of the $M1$ matrix elements to the detailed proton-neutron composition, angular momentum coupling, and spin distribution of the initial and final states.

\begin{table*}[htbp]
\centering
\caption{Comparison of theoretical and experimental \cite{nndc}  electric reduced transition probabilities for different transitions in $^{89}\mathrm{Y}$. Here, B(E2) values are presented in units of $e^2\,\mathrm{fm}^4$. Here, $J_i^{\pi}$ and $J_f^{\pi}$ are the spin-parities of the initial and final yrast states \cite{nndc}, respectively. }
\vspace{0.2cm}
\label{be2}
\resizebox{\textwidth}{!}{%
\begin{tabular}{cccccccccc}
\hline
 &  & &  &  &  &  &  & &  \\
$J_i^\pi$ & $J_f^\pi$ & Expt. & \ju & \gle-II & \gle-III & \sn-II & \sn-III & \gwb-II & \gwb-III \\
 &  & &  &  &  &  &  & &  \\
\hline
\hline
 &  & &  &  &  &  &  & &  \\
$3/2^-$ & $1/2^-$ & 49.57 (4.72) & 34.81 & 47.75 & 47.71 & 7.29 $\times 10^{-1}$ & 9.744$\times 10^{-6}$ & 1.88 & 2.00 \\

$5/2^-$ & $1/2^-$ & 54.29 (4.72) & 42.51 & 23.29 & 23.36 & 5.63 & 6.93 & 3.14 & 5.10 \\

$15/2^-$ & $13/2^-$ & 2.36 (2.12) & 29.21 & 38.29 & 38.41 & 24.38 & 34.63 & 39.72 & 47.61 \\

$17/2^-$ & $13/2^-$ & 49.57 (11.8) & 41.22 & 35.53 & 35.62 & 39.07 & 52.02 & 54.05 & 67.24 \\

$19/2^-$ & $17/2^-$ & 77.89 (49.57) & 87.19 & 80.08 & 80.08 & 85.41 & 97.90 & 89.25 & 104.10 \\

$19/2^-$ & $15/2^-$ & 51.93 (23.6) & 29.40 & 11.43 & 11.40 & 21.42 & 13.76 & 22.41 & 23.70 \\

$(21/2^-)$ & $17/2^-$ & 13.45 (3.54) & 18.00 & 37.39 & 37.41 & 43.91 & 52.33 & 43.74 & 51.32 \\

$5/2^+$ & $9/2^+$ & 101.49 (30.68) & 118.20 & 55.09 & 55.27 & 51.92 & 9.33 & 49.48 & 56.86 \\

$11/2^+$ & $9/2^+$ & 129.82 (25.96) & 141.40 & 109.00 & 109.30 & 46.33 & 178.90 & 105.90 & 212.40 \\

$13/2^+$ & $9/2^+$ & 101.49 (18.88) & 135.20 & 86.20 & 85.41 & 21.33 & 32.28  & 14.85 & 130.50 \\

$17/2^+$ & $13/2^+$ & $<$0.68 & 17.96 & 1.08 & 6.08 $\times 10^{-1}$ & 192.60 & 233.60 & 7.85 & 1.19 $\times 10^{-1}$ \\

$(23/2^+)$ & $(21/2^+)$ & 59.01 (42.49) & 107.40 & 9.03 & 86.29 & 7.94 & 130.20 & 8.34 & 124.50 \\

$(25/2^+)$ & $(21/2^+)$ & 103.85 (42.49) & 119.20 & 143.00 & 1.58 $\times 10^{-2}$  & 175.20 & $25.88 \times 10^{-2}$& 187.50 & $2.70 \times 10^{-3}$ \\

$(29/2^+)$ & $(27/2^+)$ & 188.83 (165.22) & 188.40 & 46.33 & 7.98 $\times 10^{-6}$ & 65.68 & 70.63 & 68.22 & 81.31 \\

$(31/2^+)$ & $(27/2^+)$ & $<$448.46 & 22.68 & $7.24 \times 10^{-2}$ & $9.30 \times 10^{-3}$ & 7.47 $\times 10^{-1}$ & 1.58 & 2.08 & $8.39 \times 10^{-3}$\\
 &  & &  &  &  &  &  & &  \\
 \hline
\hline
\end{tabular}%
}
\label{transition_be2}
\end{table*}


\begin{table*}[htbp]
\centering
\caption{Comparison of theoretical and experimental \cite{nndc} magnetic reduced transition probabilities for different transitions in $^{89}\mathrm{Y}$. Here, B(M1) values are presented in units of $\mu_N^2$ . Here, $J_i^{\pi}$ and $J_f^{\pi}$ are the spin-parities of the initial and final yrast states \cite{nndc}, respectively.  }
\vspace{0.2cm}
\label{bm1}
\resizebox{\textwidth}{!}{%
\begin{tabular}{cccccccccc}
\hline
 &  & &  &  &  &  &  & &  \\
$J_i^\pi$ & $J_f^\pi$ & Expt. & \ju & \gle-II & \gle-III & \sn-II & \sn-III & \gwb-II & \gwb-III \\
 &  & &  &  &  &  &  & &  \\
\hline
\hline
 &  & &  &  &  &  &  & &  \\
$3/2^-$ & $1/2^-$ & 0.47 (0.05) & 1.42 $\times 10^{0}$ & 1.45 $\times 10^{0}$ & 1.45 $\times 10^{0}$ &  $5.13 \times 10^{-1}$ &  $5.17 \times 10^{-1}$ & $3.16 \times 10^{-1}$&  $4.32 \times 10^{-1}$\\

$15/2^-$ & $13/2^-$ & 0.02 (0.00) & $6.75 \times 10^{-2}$ &  $1.93 \times 10^{-1}$ &  $1.90 \times 10^{-1}$ & $3.93 \times 10^{-1}$ & $4.66 \times 10^{-1}$&  $4.77 \times 10^{-1}$ &  $5.67\times 10^{-1}$ \\

$17/2^-$ & $15/2^-$ & 0.09 (0.02) & $3.36 \times 10^{-2}$& $2.11 \times 10^{-1}$ & $2.12 \times 10^{-1}$ & $2.40 \times 10^{-1}$ & 2.88 $\times 10^{-1}$ & 2.69 $\times 10^{-1}$ & 3.24 $\times 10^{-1}$ \\

$19/2^-$ & $17/2^-$ & 0.02 (0.01) & 1.13 $\times 10^{-4}$ & 7.52 $\times 10^{-1}$ & 7.51 $\times 10^{-1}$ & 5.83 $\times 10^{-1}$ & 7.24 $\times 10^{-1}$ & 6.51 $\times 10^{-1}$ & 8.09 $\times 10^{-1}$\\

$(21/2^-)$ & $19/2^-$ & 0.08 (0.02) & 4.86 $\times 10^{-2}$ & 7.03 $\times 10^{-1}$& 7.04 $\times 10^{-1}$ & 5.29 $\times 10^{-1}$ & $6.54 \times 10^{-1}$ & 5.78 $\times 10^{-1}$ & 6.96 $\times 10^{-1}$ \\

$11/2^+$ & $9/2^+$ & 0.00011 (0.00009) & 5.44 $\times 10^{-3}$& 1.42 $\times 10^{-2}$ & 1.42 $\times 10^{-2}$ & 5.67 $\times 10^{-3}$& 7.30 $\times 10^{-4}$ & 2.14 $\times 10^{-2}$ & 2.35 $\times 10^{-3}$ \\

$(19/2^+)$ & $17/2^+$ & $>$1.97 (0.72) & 6.90 $\times 10^{-3}$ & 8.39 $\times 10^{-3}$ & 3.04 $\times 10^{-3}$ & 2.43 $\times 10^{-2}$ & 8.41 $\times 10^{-3}$ & 1.72 $\times 10^{-2}$ & 1.74 $\times 10^{0}$ \\

$(21/2^+)$ & $(19/2^+)$ & 0.64 (0.13) & 6.95$\times$ $10^{-3}$ & 5.82$\times$ $10^{-3}$ & 1.40 $\times 10^{0}$ & 8.62$\times$ $10^{-4}$ & 6.15 $\times 10^{-1}$ & 5.58$\times$ $10^{-4}$ & 7.99 $\times 10^{-1}$ \\

$(23/2^+)$ & $(21/2^+)$ & 0.25 (0.05) & 6.14 $\times 10^{-4}$ & 7.84 $\times 10^{-3}$ & 3.70 $\times 10^{-1}$ & 7.95 $\times 10^{-3}$ & 9.73 $\times 10^{-1}$ & 1.38 $\times 10^{-2}$ & 8.51 $\times 10^{-1}$ \\

$(25/2^+)$ & $(23/2^+)$ & 0.04 (0.01) & 3.74 $\times 10^{-4}$ & 9.48 $\times 10^{-3}$ & $3.92 \times 10^{-6}$ & 9.09 $\times 10^{-3}$ & $4.39 \times 10^{-5}$ & 7.80 $\times 10^{-3}$ & $1.46 \times 10^{-8}$ \\

$(27/2^+)$ & $(25/2^+)$ & 0.11 (0.03) & 1.42 $\times 10^{-3}$ & 2.08 $\times 10^{-2}$ & $2.23 \times 10^{-5}$ & 5.45 $\times 10^{-3}$ & 1.42 $\times 10^{-2}$ & 6.08 $\times 10^{-3}$ & 1.69 $\times 10^{-2}$ \\

$(29/2^+)$ & $(27/2^+)$ & 0.48 (0.16) & 1.56 $\times 10^{-1}$ & 9.34 $\times 10^{-1}$ & 3.95 $\times$ $10^{-10}$ & 7.37 $\times 10^{-1}$ & 8.86 $\times 10^{-1}$ & 8.04 $\times 10^{-1}$ & 1.047 $\times 10^{0}$ \\

$(31/2^+)$ & $(29/2^+)$ & 0.77 (0.30) & 1.94 $\times 10^{-1}$ & $6.489 \times 10^{-6}$ & $1.13 \times 10^{-4}$ &  $3.93 \times 10^{-4}$&  $1.96 \times 10^{-3}$& $2.52 \times 10^{-4}$&  $2.30 \times 10^{-3}$\\
 &  & &  &  &  &  &  & &  \\
\hline
\end{tabular}%
}
\label{transition_bm1}
\end{table*}

\section{\label{Summary} Summary and Conclusion}
The systematic shell-model study of \(^{89}\mathrm{Y}\) using different effective interactions and different neutron cross shell truncations reveals a clear evolution of its microscopic structure with increasing excitation energy and angular momentum. The low-lying states are predominantly governed by proton excitations within the \(f_{5/2}pg_{9/2}\) model space, accounting for the reasonable description obtained with the restricted \(\ju\) interaction. With increasing spin. However, proton excitations alone become insufficient and neutron excitations across the \(N=50\) shell gap play an increasingly important role in describing high spin states. In particular, the relatively low-lying \(\nu2d_{5/2}\) orbital provides a favorable pathway for neutron core excitation and substantially improves the description of the intermediate- and high spin spectrum. Also, excitation to the higher-lying \(\nu1g_{7/2}\) orbital is comparatively suppressed. Occupation-number analyses further demonstrate a gradual transition from proton dominated configurations to states involving significant neutron core excitation, with proton pair breaking and neutron excitation contributing on comparable mechanism at higher spins. The electromagnetic transition strengths  \(B(M1)\),  \(B(E2)\) demonstrate the increasing sensitivity of the higher-spin structure to proton-neutron correlations and neutron cross shell excitations. While the importance of neutron core excitations in the high spin structure of \(^{89}\mathrm{Y}\) has been established in earlier studies, the present work provides a systematic comparison of different cross shell interactions and truncation schemes and traces the evolution of the proton and neutron contributions with angular momentum. In particular, the combined analysis of excitation energies, occupation probabilities, \(I_p\otimes I_n\) decomposition, and electromagnetic transition strengths identifies the \(\nu2d_{5/2}\) excitation as a particularly important component of the high spin structure. Overall, the present results indicate that the structure of \(^{89}\mathrm{Y}\) evolves from predominantly proton dominated configurations at low spin to strongly competing and mixed proton-neutron configurations at high spin.

\section*{Acknowledgment}
The authors acknowledge the IEDC Laboratory, Department of Basic Science and Humanities, IEM Kolkata (Newtown Sector), for providing computational facilities.

\section*{Roles}
\textbf{Sangeeta Das:} Conceptualisation,  computation, coding, data analysis, plotting, and manuscript writing. \textbf{Arkabrata Gupta} Computation, coding, data analysis, plotting, and manuscript writing.

\bibliography{Reference}

\end{document}